\documentclass[12pt]{article}
\usepackage{amsmath}
\usepackage{graphicx,psfrag,epsf}
\usepackage{enumerate}
\usepackage{natbib}
\usepackage{hyperref}
\usepackage{authblk}
\usepackage{hypernat}
\usepackage{url}
\usepackage{subcaption}
\usepackage{amsfonts}
\usepackage{amsmath,bm, amssymb}
\usepackage{mathtools}

\hypersetup{breaklinks=true,
            pdfauthor={Devin Francom (Los Alamos National Laboratory), J. Derek Tucker (Sandia National Laboratories), Gabriel Huerta (Sandia National Laboratories), Kurtis Shuler(Sandia National Laboratories), and Daniel Ries (Sandia National Laboratories)},
            pdfkeywords = {Bayesian inference, model calibration},
            pdftitle={Elastic Bayesian Model Calibration},
            colorlinks=true,
            citecolor=blue,
            urlcolor=blue,
            linkcolor=magenta,
            pdfborder={0 0 0}}

\DeclareMathOperator*{\arginf}{arg\,inf}

\DeclareMathOperator{\sgn}{sgn}

\newcommand{\s}{\ensuremath{\mathbb{S}}}

\newcommand{\ltwo}{\ensuremath{\mathbb{L}^2}}

\usepackage{xcolor}
\definecolor{Blue}{RGB}{0,119,255}

\title{Elastic Bayesian Model Calibration}
\author[1]{Devin Francom}
        \author[2]{J. Derek Tucker}
        \author[2]{Gabriel Huerta}
        \author[2]{Kurtis Shuler}
        \author[2]{Daniel Ries}
        \affil[1]{Los Alamos National Laboratory}
        \affil[2]{Sandia National Laboratories}

\date{\today}

\begin{document}

\maketitle
	
	\begin{abstract}  
        Functional data are ubiquitous in scientific modeling.  For instance, quantities of interest are modeled as functions of time, space, energy, density, etc.  Uncertainty quantification methods for computer models with functional response have resulted in tools for emulation, sensitivity analysis, and calibration that are widely used.  However, many of these tools do not perform well when the computer model's parameters control both the amplitude variation of the functional output and its alignment (or phase variation). This paper introduces a framework for Bayesian model calibration when the model responses are misaligned functional data.  The approach generates two types of data out of the misaligned functional responses: (1) aligned functions so that the amplitude variation is isolated and (2) warping functions that isolate the phase variation.  These two types of data are created for the computer simulation data (both of which may be emulated) and the experimental data. The calibration approach uses both types so that it seeks to match both the amplitude and phase of the experimental data.  The framework is careful to respect constraints that arise especially when modeling phase variation, and is framed in a way that it can be done with readily available calibration software.  We demonstrate the techniques on two simulated data examples and on two dynamic material science problems: a strength model calibration using flyer plate experiments and an equation of state model calibration using experiments performed on the Sandia National Laboratories' Z-machine.
	\end{abstract}
    
\noindent%
    \vfill

\section{Introduction} 
In domains of science and engineering where modeling is an important part of investigation and discovery, quantifying uncertainty in model inferences and predictions can be essential in order for model performance to be trusted.  When a model has uncertain parameters (or inputs), model calibration is the act of tuning the parameters so that the model produces a desired response.  Most often, models are calibrated to experimental or observational measurements, so that calibration seeks to make the model response reflect reality. Model calibration (sometimes known as inversion) is often a poorly identified problem, where multiple combinations of inputs produce equally valid solutions.  \cite{kennedy2001bayesian} proposed Bayesian model calibration as a systematic approach to calibrating a model in the face of all of the sources of uncertainty, so that calibration uncertainty is quantified.  These sources of uncertainty are parameter uncertainty, measurement uncertainty, emulation or surrogate model uncertainty (an emulator is a fast surrogate for a more expensive model), and model form error (i.e., simulation model misspecification, discrepancy, or bias).  

Often, computer model outputs are functional in nature, producing an output measured over space and/or time, for example. The majority of calibration work on these types of outputs has been done on features extracted from the outputs such as peak points or critical values \citep{walters2018bayesian}. 
However, the process for extracting these features can be tedious, prone to error, and is problem specific.  Other calibration approaches have been developed for use with multivariate or functional response \citep{higdon2008computer,bayarri2007AS,francom2019inferring}, though they are prone to problems when presented with functional response data that are misaligned.
Our goal is to extend functional emulation and calibration methods for use when the response is misaligned, and to do so without human-intensive feature engineering.

\begin{figure}
    \centering 
    \includegraphics[width=\textwidth]{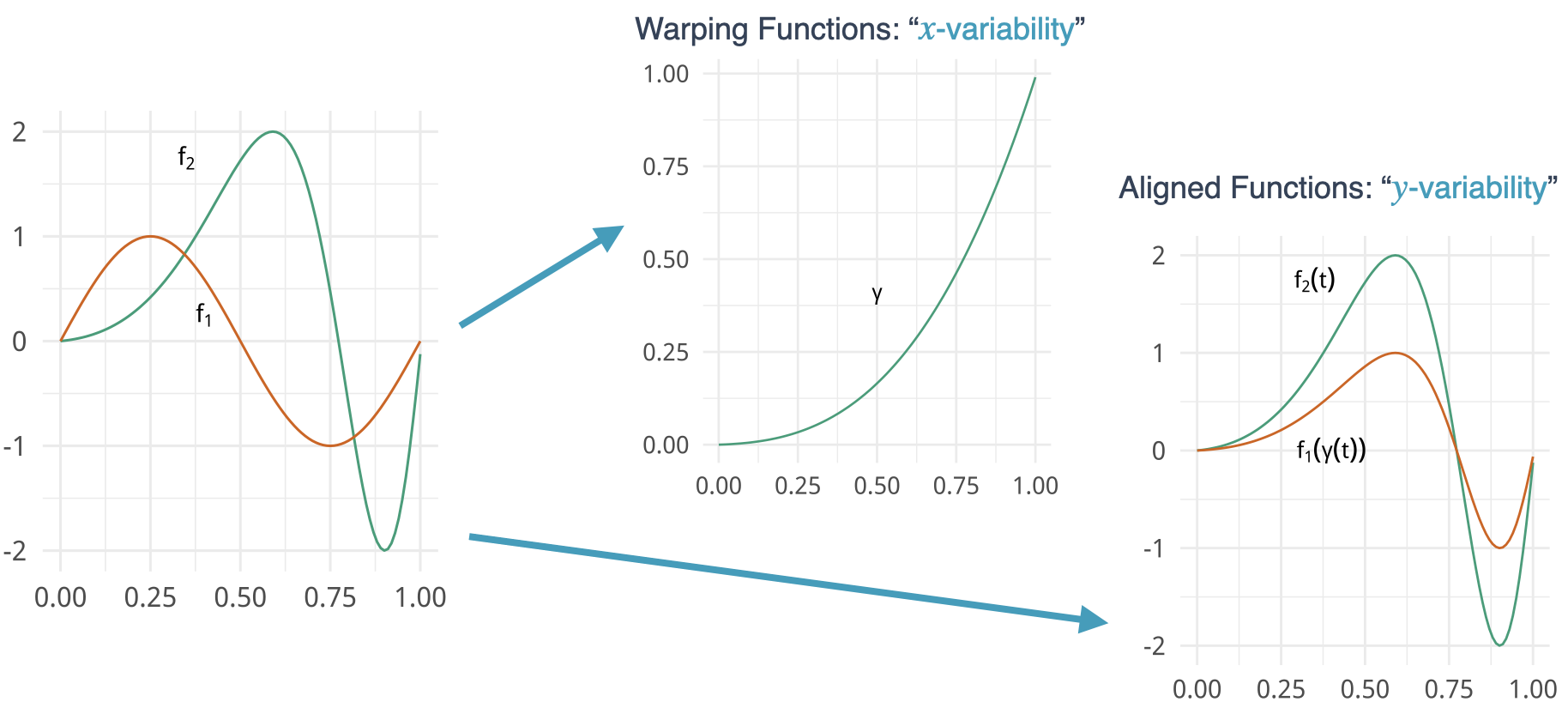}  

    \caption{Demonstration of amplitude and phase variability in functional data. Left: Original functions. Middle: Warping Function ("phase variability"). Right: Aligned functions ("amplitude variability").}
    \label{fig:toy_example}
\end{figure}

There has been considerable effort in statistics to
develop methods that can analyze functional data objects without loss of
information. Such methodology is known as functional data analysis and
has a rich history. An excellent introduction to this
field is given in several books including \cite{ramsay2005},
\cite{horvath2012}, and \cite{srivastava2016}. An interesting
aspect of most functional data is that the underlying variability can be
ascribed to two sources. These two sources are termed the amplitude (\(y\) or vertical) variability and the phase (\(x\) or horizontal or
warping) variability. Capturing these two sources of variability is
crucial when modeling functional data, and can greatly affect the construction of
statistics (e.g., averages, tolerance bounds). 
In this work, we refer to methods that handle both amplitude and phase variability in functional data as \emph{elastic}. This
important concept is illustrated in Figure \ref{fig:toy_example} through
a simulated example. 

In Figure \ref{fig:toy_example} we have two functions that contain both a peak and a trough. Each of the functions contain variability in the height of the peak and valley and large variability in its placement. 
The relative heights of the peaks can be attributed to the amplitude
variability, while the different locations of the peaks constitute the
phase variability.

The phase variability can be accounted for by first
aligning the functions. As an example, the right panel shows
time-aligned functions. The alignment involves a transformation of the
horizontal axis via the warping function shown in the middle panel.
The aligned function ($f_1(\gamma(t))$) captures the amplitude variability while the
warping function ($\gamma$) captures the phase variability. The
Bayesian model calibration method introduced in this paper considers the
shape of the data by accounting for both directions of variability.

While standard calibration solutions can be applied to misaligned functions especially when emulation is unnecessary, we demonstrate that using functional metrics that consider the misaligned nature of the data can produce more accurate calibration solutions and more practical model formulations.
In our methodology, we use elastic functional data analysis methods \citep{srivastava2016,tucker-wu-srivastava:2013,marron2015} to construct a metric to measure the distance between functions and to specify a likelihood. 
We do this by decomposing our functional responses into aligned functions and warping functions, as shown in Figure \ref{fig:toy_example}. 
For expensive models we can then build an emulator or surrogate for the aligned functional responses, and, under a suitable transformation, an emulator for the warping functions.  These two emulators can be used to calibrate in such a way that a proper distance is used.  The strengths of this approach are that (1) emulation is likely to be more accurate when applied to aligned data instead of misaligned data \citep{francom2022landmark}, (2) discrepancy and measurement error modeling can be done in such a way that it is isolated to the phase or the amplitude part of the model, and (3) the alignment procedure ensures the isometry property holds and therefore, the calibration is performed using a proper distance (i.e., Euclidean distance in the transformed space properly reflects distance between warping functions). Hence, this work overtly connects the two well-established fields of Bayesian model calibration and functional data analysis that have historically had limited connection. This enables better emulation, more identifiable discrepancy modeling, and hence more accurate (likely smaller) calibration uncertainty. Further, we do this in such a way that it is accessible using established tools, not requiring new and specialized or expensive approaches to posterior sampling. The misalignment problem has not been adequately addressed (or even identified) in the Bayesian model calibration literature, and our method produces significantly better results than approaches that ignore the misalignment or only partially deal with it.
  Related work includes \cite{kleiber2014model}, which proposed a functional calibration approach using a deformation that relies on the $\ltwo$ metric in the standard function space. However, this metric can exhibit a ``pinching effect'' degeneracy which the elastic methods do not have, as demonstrated in \cite{srivastava2016}. Recent work by \cite{Guan2019} performed Bayesian model calibration utilizing phase-variability, however the emulation was done on the likelihood, limiting extension to the richer class of calibration models that we consider (e.g., including discrepancy and measurement error). 
We demonstrate our approach on two dynamic material science applications described in the following subsections.

\subsection{Equation of state exploration using Z-machine experiments}

\begin{figure}[htbp]
\centering
\includegraphics[width=0.9\textwidth]{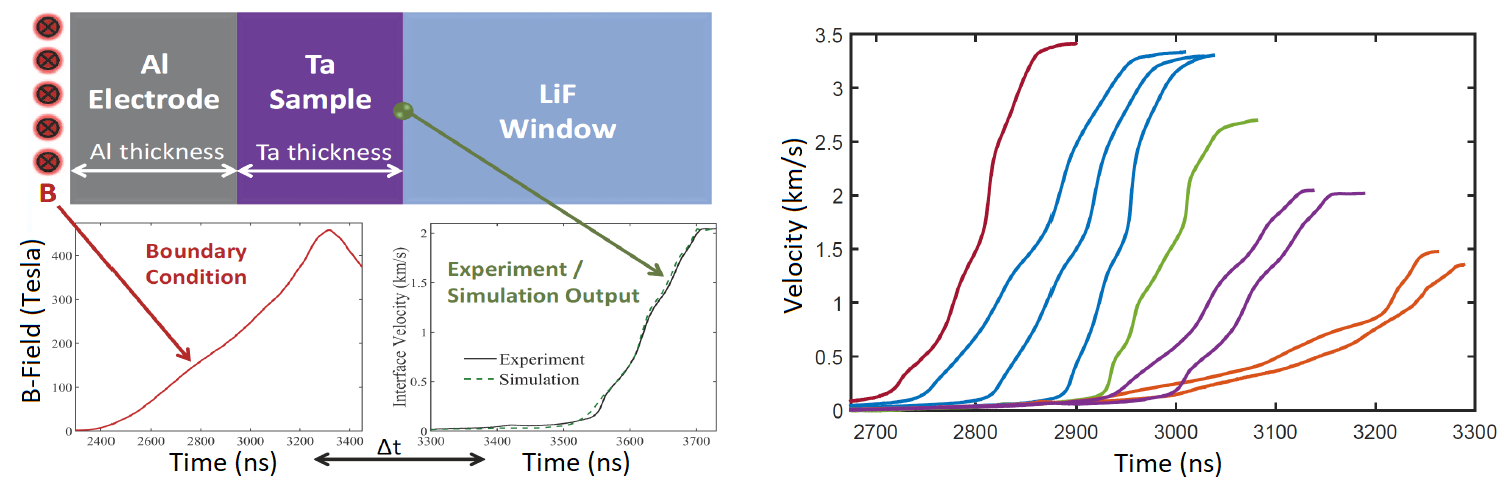}
\caption{From \cite{brown2014} and \cite{brown2018}. Left: Example of experiments at the Z-machine. Right: Velocity response functions from nine experiments. }
\label{fig:ta_data2}
\end{figure}

This problem is motivated by data from \cite{brown2014} which conducted dynamic materials experiments on tantalum (Ta) generated with pulsed magnetic fields using Sandia National Laboratories' Z-machine \citep{savage2007}. A basic version of the experiments performed at the Z-machine can be seen in Figure \ref{fig:ta_data2} (left). 
The Z-machine delivers electrical currents along an aluminum (Al) panel creating massive pressure driving an impulse into a tantalum sample. 
Stress waves flow through the tantalum, and the experiment results in measurements of velocity on the far side of the sample as a function of time. More details on the execution of the experiment can be found in \cite{brown2014} and \cite{lemke2005}. Figure \ref{fig:ta_data2} (right) shows the functional response of nine experiments.

The computer model input parameters consist of calibration parameters of key interest to an equation of state (EoS) of tantalum and experiment-specific parameters that are not completely known. More details about these input parameters appear in  \ref{sec:z-machine}.

\subsection{Material strength exploration using gas gun experiments}

\begin{figure}[htbp]
    \centering
    \includegraphics[width=0.6\textwidth]{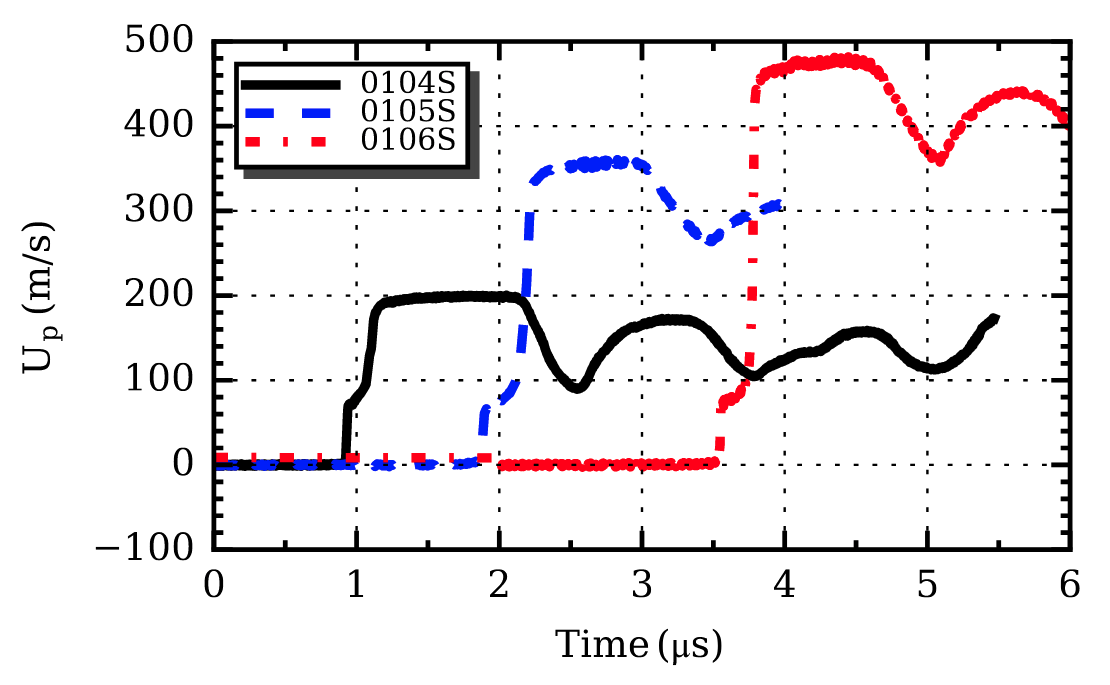}
    \caption{From \cite{walters2018bayesian}, the response from three gas gun experiments where both flyer and plate were aluminum alloys.}
    \label{fig:walters}
\end{figure}

\cite{boteler2006dynamic} performed a series of experiments where a flyer was shot from a gas gun into a plate and the velocity of the opposite surface of the plate was measured (as a function of time) as the resultant shock wave moved through it.  This experiment has similarities to the Z-machine example above, except that the shock waves that propagate through the plate are driven by the flyer's impact instead of the more continuous electrical current drive.  Figure \ref{fig:walters} shows three velocimetry curves measured during three of these flyer plate gas gun experiments.  \cite{walters2018bayesian} used these experiments to parameterize a strength model for aluminum. Also similar to the Z-machine computer model, flyer plate impact simulations include both inputs to be calibrated (strength model parameters) and experiment-specific parameters that are uncertain (e.g., the exact velocity of the flyer).

The structure of the rest of this paper is as follows. Section \ref{sec:review} gives a high level overview of model calibration based on the  \cite{kennedy2001bayesian} framework and functional response extensions. 
Section \ref{sec:align} describes the types of variability for functional output, functional alignment with a proper distance metric, and how to measure the amount of variability in amplitude and phase space. 
Section \ref{sec:ebcalibration} describes how to use the measures of variation in our elastic Bayesian model calibration method, and various modeling choices are described in Section \ref{sec:ebmc_choices}.
Section \ref{sec:simresults} gives two simulated examples comparing the proposed elastic functional calibration method with functional response calibration that does not account for misalignment.
Section \ref{sec:results} applies the proposed method to the two material model calibration problems detailed above. 
Lastly, Section \ref{sec:discussion} provides general conclusions and considerations on future work. Supplementary material includes an additional simulated example.

\section{Bayesian Model Calibration}
\label{sec:review}
\subsection{Univariate Response}

The traditional approach to Bayesian model calibration, introduced by \cite{kennedy2001bayesian} and expanded in \cite{higdon2004combining}, seeks to calibrate parameters of a model using observations. Let $y(\bm x,\bm u)$ denote the model where $\bm x$ denotes conditions that are certain, often fixed conditions of an experiment, and $\bm u$ denotes uncertain parameters in need of calibration. Let $z(\bm x)$ denote an observation.  In these approaches, $\bm x$ could include functional variables like space or time and/or conditions at which an experiment was performed.  Let $n$ be the number of observations or experimental conditions measured, so that our observed data are $z(\bm x_1),\dots,z(\bm x_n)$.  Then the calibration model is
\begin{align}
z(\bm x_i) &= y(\bm x_i,\bm \theta) + \delta(\bm x_i) + \epsilon(\bm x_i),~~ \epsilon(\bm x_i) \sim \mathcal{N}(0,\sigma_\epsilon^2)\label{eq:koh1}
\end{align}
where $\bm\theta$ denotes the best set of calibration parameters for $\bm u$, $\delta$ denotes (latent) error in the form of the model (often called model discrepancy), and $\epsilon$ denotes measurement or observation error in $z$. The Gaussian likelihood specified here assumes that measurement errors are independent and identically distributed, but other measurement error structures can be used.  The unknowns in this model are the calibration parameters $\bm \theta$, the measurement error variance $\sigma^2_\epsilon$, and the form of the discrepancy function.  In order for this model to be identifiable, priors for each of the unknowns need to be chosen carefully.  If $n$ is small, the prior for $\sigma^2_\epsilon$ will be influential.  Additionally, there is a natural trade off between the calibration parameters and the model discrepancy, so that at least one of these needs to be well constrained by the prior.  For instance, we could use a Gaussian process prior for the discrepancy function $\delta$ in such a way that we prefer small-valued smooth functions \citep{higdon2004combining}, or we could try to specify priors that prefer positive values or monotone functions \citep{bryn2014}.

In many realistic scenarios the model $y$ is expensive to evaluate, requiring powerful computing and non-trivial amounts of time.  This slows down the evaluation of the likelihood and makes inference impractical.  In these scenarios, we build a surrogate model (or emulator) to use in place of $y$, which is trained using a (relatively) small number of model evaluations, $\{y(\bm x_j, \bm u_j)\}_{j=1}^{N_{sim}}$ \citep{sacks1989SS}.  The full Bayesian approach to inference then seeks to learn all the unknowns (calibration parameters, model discrepancy, measurement error, and emulator parameters) conditional on all of the data ($n$ observations and $N_{sim}$ model runs).  \cite{higdon2004combining} use this approach with a Gaussian process emulator, while \cite{kennedy2001bayesian} fix some of the Gaussian process emulator and/or discrepancy parameters in advance for computational and identifiability reasons. More generally, \cite{liu2009modularization} describe cases where modularization (or {\it cutting feedback} from parts of the model that lead to identifiability issues or poor MCMC mixing) will result in better posterior inference. 
Many modern practitioners use modularized approaches either for philosophical or computational reasons.

\subsection{Functional Response}
\label{sec:pointwise_cal}
Various research efforts have extended the Bayesian model calibration approach above to be more explicitly suited to functional response.  This can vastly increase the training data size, but Gaussian processes with Kronecker covariance structures can help produce scalable models \citep{williams2006combining}.  A somewhat different approach is used by \cite{gu2016parallel}, which involves fitting a Gaussian process for each output while sharing some parameters across processes, and, with a suitable discrepancy model, the calibration can be obtained. 
Perhaps the most common approach is to project the functional response onto basis functions and build functional models for calibration in the reduced dimension basis coefficient space \citep{higdon2008computer,bayarri2007AS,francom2019inferring}. Let $t$ denote the functional variable (e.g., time), and let $z(t,\bm x_i)$ denote an experimental measurement from the $i^{th}$ experiment at functional variable $t$.  Similarly, let $y(t,\bm x_i, \bm u)$ denote a simulation of the $i^{th}$ experiment at the functional variable $t$ with input parameters $\bm u$.  Then an approach to Bayesian model calibration with functional response specifies
\begin{align}
    z(t,\bm x_i) &= y(t,\bm x_i,\bm \theta) + \delta(t, \bm x_i) + \epsilon(t,\bm x_i), ~~\epsilon(t,\bm x_i)\sim \mathcal{N}(0,\sigma_\epsilon^2) \label{eq:fbmc}.
\end{align}
The typical approach to inference is to discretize $t$ onto a grid $t_1,\dots,t_{N_T}$, which generates $N_T$ dimensional vectors $\bm z(\bm x_i)$, $\bm y(\bm x_i,\bm\theta)$, and $\bm\delta(\bm x_i)$ of the respective functions evaluated on the discretized grid, which simplifies the model to a multivariate representation,
\begin{align}
    \bm z(\bm x_i) &= \bm y(\bm x_i,\bm\theta) + \bm\delta(\bm x_i) + \bm\epsilon(\bm x_i), ~~\bm\epsilon(\bm x_i)\sim N(\bm 0,\sigma_\epsilon^2 \bm I).
\end{align}
 As in the univariate case, when $y(t,\bm x, \bm \theta)$ is expensive to evaluate, we will require an emulator or surrogate model in order to evaluate the likelihood function quickly.  \cite{higdon2008computer} projects the model runs onto functional principal components and projects the discrepancy onto a separate flexible basis, and inference is carried out in the resulting low dimensional space. \cite{bayarri2007framework} project the model runs and discrepancy onto a wavelet basis and carry out the inference in the low dimensional space.  \cite{francom2019inferring} projects the model runs onto functional principal components, allows the discrepancy to be full-dimensional, and fits the emulator in a modular fashion by keeping aspects of the emulator separate from the full Bayesian analysis.

All of these calibration approaches can be applied to misaligned functional data, but all will have difficulty with emulation and the specification of model discrepancy, as they will only deal explicitly with amplitude variation.  Additionally, distances computed between the different functions will not be proper distances since they would not account for the phase variability.

\section{Elastic Bayesian Model Calibration}
\label{sec:functionalcal}
In this section, we will review elastic functional data analysis, introduce our elastic approach to calibration, and suggest various modeling choices and practical considerations.

\subsection{Elastic Functional Data Analysis}
 \label{sec:align}

\subsubsection{Types of variability in functional outputs}

When model outputs are functional, we must consider two types of variability in the functional outputs: amplitude and phase variability.  Amplitude variability is variability in the output for a fixed time ($t$), or, more simply, $y$-axis variation.  Phase variability is variability in time, or, more simply, $x$-axis variability. 

For computer model applications, input variables can induce both phase and amplitude variability in the computer model outputs, resulting in fundamentally different shapes over the range of plausible inputs.  
Additionally, model discrepancy can result in an imperfect match between the computer model prediction and observed data at the correct value of the model inputs. With misaligned functional data, we must consider model discrepancy in both phase and amplitude variability to accurately represent discrepancy-induced shape distortions in the functional predictions. 

When functional response variability is not solely driven by amplitude variability, pointwise (non-elastic) calibration (as in Section \ref{sec:pointwise_cal}) will produce  estimates of the calibration parameters that may not agree with the true physical values, when these exist. Even when phase variability is taken into account, there exists a problem with the $\ltwo$ metric known as the \textit{pinching problem} \citep{ramsay-li:1998}. Specifically, if we have two
functions, $f_1$ and $f_2$ and the range$(f_1)$ is entirely above the
range$(f_2)$, the $\ltwo$ metric becomes degenerate and pinches the warped
function. To address this problem, \cite{srivastava-etal-JASA:2011} introduced a mapping for functional data 
called the {\it square-root velocity function} or SRVF 
that improves functional alignment, and provides fundamental
mathematical equalities that lead to the formal development of this topic.
Moreover, the metric used in the alignment is a proper distance and avoids the
pinching effects of the standard $\ltwo$ metric in function space without the use of a penalty.  We propose functional calibration metrics that account for both phase and amplitude variation while properly measuring the distance between functions.

\subsubsection{Functional Alignment}

To explain metrics for comparing functional data in a calibration setting, we consider the comparison of two functions of $t$: $z(t,\bm x_i)$ and $y(t,\bm x_i, \bm \theta)$.  Varying $\bm\theta$ will change the shape of $y(t,\bm x_i, \bm \theta)$ so we seek to find $\bm\theta$ such that $z(t,\bm x_i)$ and $y(t,\bm x_i, \bm \theta)$ are \textit{optimally matched}, where the optimality criterion considers the distance between the functions in both amplitude and phase.  For notational convenience, and to emphasize that we are comparing these functions for a fixed $\bm x_i$ and $\bm\theta$, we rewrite $z(t,\bm x_i)$ and $y(t,\bm x_i, \bm \theta)$ as $z(t)$ and $y(t)$ for the rest of this subsection.  To measure the distance between $z(t)$ and $y(t)$, we use elastic functional data analysis (EFDA) \citep{srivastava2016}.  The main premise behind EFDA is to construct a proper distance metric between the computational prediction $y(t)$ and the experimental data $z(t)$.  To construct this metric, a continuous mapping $\gamma_{y\rightarrow z}(t): [0,1] \rightarrow [0,1]$ between $y(t)$ and the experimental data $z(t)$ is constructed such that $\gamma$ is a diffeomorphism. The function $\gamma_{y\rightarrow z}(t)$ is referred to as a warping function, as it measures phase distortions in $y(t)$ such that $y\circ \gamma_{y\rightarrow z}(t) = y(\gamma_{y\rightarrow z}(t))$ aligns with $z(t)$.  \cite{srivastava-etal-JASA:2011,tucker-wu-srivastava:2013} show that by applying a specific transformation to the original functions $z(t)$ and $y(t)$, there exist simple expressions for the amplitude and phase distance between functions.  We now describe how to construct this transformation and how to define the distance metrics on the transformed data.  

The functions $z(t)$ and $y(t)$ are transformed to their square-root velocity functions (SRVFs).  
That is, we define the SRVF of $f(t)$ as:
\begin{align}
q(t) &= \sgn(\dot{f}(t))\sqrt{|\dot{f}(t)|}
\end{align}
where $\dot{f}$ denotes the time derivative of $f$.  
The SRVF is a bijective mapping up to a translation; that is, $f(t)$ can be uniquely determined from $q(t)$ and a single point on the curve $f(t)$.

Let $q_z(t)$ and $q_y(t)$ denote the SRVFs of $z(t)$ and $y(t)$ respectively.  The warping function that aligns $y$ to $z$, denoted $\gamma_{y\rightarrow z}$, can be estimated solving the following optimization problem via Dynamic Programming \citep{tucker-wu-srivastava:2013}
\begin{align}
\label{eq:optim}
\gamma_{y\rightarrow z} = \arginf_{\gamma\in\Gamma} || q_z - (q_y \circ \gamma) \sqrt{\dot{\gamma}} ||^2
\end{align}
or using a Bayesian approach \citep{cheng2016bayesian, lu2017bayesian}. We utilize the same group structure as \citep{tucker-wu-srivastava:2013} where $\Gamma$ is the set of orientation-preserving diffeomorphisms of the unit interval $[0,1]$: $\Gamma = \{\gamma: [0,1] \rightarrow [0,1] |~\gamma(0) = 0,~\gamma(1)=1,\gamma~\textnormal{is a diffeomorphism} \}.$ 
The properties of the diffeomorphisms is what allows the bending and stretching described above and why we use the term \textit{elastic} and the norm is the standard $\ltwo$ norm on the space of SRVFs.

An advantage of this approach to warping function estimation is that the analyst does not have to specify landmarks for function alignment; the estimation of $\gamma$ is achieved by using the group structure of $\Gamma$.  When we use the optimization in Equation \ref{eq:optim} in later sections, we will refer to this as a decomposition (specifically the \textit{warping decomposition}) because it decomposes a misaligned function into an aligned function and a warping function.

Given a warping function,
$\gamma(t)$, and aligned SRVF, 
$(q_y \circ \gamma \sqrt{\dot{\gamma}})(t)$, we can construct measures of phase and amplitude variability.  Specifically, amplitude variability is measured as:
\begin{align}
	d_a(q_z,q_y) &= ||q_z - (q_y\circ \gamma_{y\rightarrow z})\sqrt{\dot{\gamma}_{y\rightarrow z}}||^2,
\end{align}
\cite{srivastava-etal-JASA:2011} show that this $\ltwo$ distance on the transformed and aligned SRVF functions is a proper distance metric for amplitude.

\subsubsection{Measuring phase distance} 

Defining a measure of phase variability is more difficult than for amplitude variability, because the space of warping functions, $\Gamma$, is an infinite-dimensional nonlinear manifold, and therefore cannot be treated as a standard Hilbert space.  
Since we would like to exploit Riemannian-geometric structure when making inferences about the warping functions, we again apply a specific transformation to the warping functions such that we can use a standard distance metric (the $\ltwo$ norm) to measure distance. 
Specifically, we represent an element $\gamma \in \Gamma$ by the square-root of its derivative $\psi = \sqrt{\dot{\gamma}}$. Note that this is the same as the SRVF defined earlier, and takes this simpler form because $\dot{\gamma} > 0$. The identity $\gamma_{z\rightarrow z}$ maps to a constant function with value $\psi_{z\rightarrow z}(t) = 1$, which corresponds to no warping. Since $\gamma(0) = 0$, the mapping from $\gamma$ to $\psi$ is a bijection and one can reconstruct $\gamma$ from $\psi$ using $\gamma(t) = \int_0^t \psi(s)^2 ds$.  
An important advantage of this transformation is that since $\| \psi\|^2 = \int_0^1 \psi(t)^2 dt = \int_0^1 \dot{\gamma}(t) dt = \gamma(1) - \gamma(0) = 1$, the set of all such $\psi$'s is the positive orthant of the Hilbert sphere $\Psi=\s_{\infty}^+$ (i.e., a unit sphere in the Hilbert space $\ltwo$). 
In other words, the square-root representation simplifies the complicated geometry of $\Gamma$ to a unit sphere. The distance between any two warping functions, i.e., the phase distance, is exactly the arc-length between their corresponding SRVFs on the unit sphere
$\s_{\infty}$: \[
d_{p}(\gamma_1, \gamma_2) = d_{\psi}(\psi_1, \psi_2) \equiv \cos^{-1}\left(\int_0^1 \psi_1(t) \psi_2(t) dt \right)\ .\]

While the geometry of $\Psi\subset\s_{\infty}$ is more tractable, it is still a nonlinear manifold and computing distances remains
difficult. Instead, we use a tangent (vector) space at a certain fixed
point for further analysis. The tangent space at any point
$\psi \in \Psi$ is given by:
$T_{\psi}(\Psi) = \{v \in \ltwo| \int_0^1 v(t) \psi(t) dt = 0\}$. To
map between the representation space $\Psi$ and tangent spaces, one
requires the exponential and inverse-exponential mappings. The
exponential map at a point $\psi\in\Psi$ denoted by
$\exp_\psi : T_{\psi}(\Psi) \mapsto \Psi$, is defined as

\begin{align}
\exp_\psi(v) &= \cos(\|v\|)\psi+\sin(\|v\|)\frac{v}{\|v\|},
\end{align}

\noindent where $v\in T_{\psi}(\Psi)$. Thus, $\exp_\psi(v)$ maps
points from the tangent space at $\psi$ to the representation space
$\Psi$. Similarly, the inverse-exponential map, denoted by
$\exp_{\psi}^{-1} : \Psi \mapsto T_{\psi}(\Psi)$, is defined as

\begin{align}
\exp_{\psi}^{-1}(\psi_1) &= \frac{\kappa}{\sin(\kappa)}(\psi_1-\cos(\kappa)\psi),
\end{align}
\noindent where $\kappa = d_{p}(\gamma_1, \gamma)$. This mapping takes
points from the representation space to the tangent space at $\psi$.

The tangent space representation $v$ is sometimes referred to as a
\emph{shooting vector}. As discussed previously, it can be sensible to warp simulations to the observations in this calibration framework, which corresponds to defining the
tangent space relative to the observations $z$, which we denote as $\psi_{z\rightarrow z}$. The identity warping is defined as $\gamma_{z\rightarrow z}(t)=t$, which results in $\psi_{z\rightarrow z}(t)=1$ and $\exp_{\psi_{z\rightarrow z}}^{-1}(\psi_{z\rightarrow z}(t))=0$.
In this case, deviations in $v$ from 0 represent deviation in phase from the experimental data.

In practice, there are situations when warping to the experiment presents difficulties, which we discuss in a later section. 
In these cases, we warp model runs and experiments to a common ``template'' function (e.g., one of the model runs that has all the peaks/valleys/characteristics of interest, possibly selected by sub-screening the model runs visually). 
We then define distance between a model run and the experiment by considering their distances to the template. 
This is effectively changing the origin in the distance calculation. 
Just as changing the origin when calculating Euclidean distances does not change the distance between two points, because the metric defined above is a proper distance metric, mathematical distance between misaligned curves is unaffected by the choice of template. 
In practice there can be differences, which leads us to introduce elastic Bayesian model calibration in the next section using an arbitrary template, $y^*$, typically taken to be the experiment or one of the model runs. 
Hence, discussion of warping in the next section will involve warping experiment and simulations to the template to get quantities like $\gamma_{y\rightarrow y^*}$ and $\gamma_{z\rightarrow y^*}$ instead of the quantities used above, $\gamma_{y\rightarrow z}$ and $\gamma_{z\rightarrow z}$.

\subsection{Using Elastic FDA within Bayesian Model Calibration}\label{sec:ebcalibration}
With the metrics for distance defined in the previous section, we are ready to detail how to do Bayesian model calibration with misaligned functional responses, which we call elastic Bayesian model calibration.

We decompose the observations from experiment $i$ into aligned functions and warping functions (each specific to experiment $i$) so that
\begin{align}
    z(t,\bm x_i) &= \tilde z(t, \bm x_i) 
    \circ_t\gamma_{z\rightarrow y^*}(t,\bm x_i)
\end{align}
where $\circ_t$ emphasizes that the composition is only in $t$, such that $f(a,b) \circ_a g(a,b) = f(g(a,b),b)$.  If the template $y^*$ is taken to be the observations $z$, this is the identity warping and nothing happens in this step.
We similarly decompose each simulation (e.g., simulation $j$) with
\begin{align}
    y(t,\bm x_j, \bm u_j) &= \tilde y(t,\bm x_j, \bm u_j) \circ_t \gamma_{y\rightarrow y^*}(t,\bm x_j, \bm u_j).
\end{align}
using the warping decomposition of Equation \ref{eq:optim}.
To facilitate modeling with proper distance metrics, we transform the warping functions into shooting vector space with
\begin{align}
    \bm v_{z\rightarrow y^*}(\bm x_i) &= \exp_{\psi_{z\rightarrow y^*}}^{-1}\left(\sqrt{\dot\gamma_{z\rightarrow y^*}(\bm x_i)}\right) \\
    \bm v_{y\rightarrow y^*}(\bm x_j,\bm u_j) &= \exp_{\psi_{y\rightarrow y^*}}^{-1}\left(\sqrt{\dot\gamma_{y\rightarrow y^*}(\bm x_j,\bm u_j)}\right) .
\end{align} 
We can then use these aligned model runs and associated shooting vectors for emulator building, if desired. 

\subsubsection{Calibration when Emulation is Unnecessary}
Our calibration model using the aligned data and shooting vectors is specified with
\begin{align}
    \tilde z(t,\bm x_i) &= \tilde y(t,\bm x_i, \bm\theta) + \delta_{\tilde y}(t,\bm x_i) + \epsilon_{\tilde z}(t,\bm x_i), ~~\epsilon_{\tilde z}(t,\bm x_i)\sim \mathcal{N}(0,\sigma_{\tilde z}^2) \label{eqn:ye}\\
    \bm v_{z\rightarrow y^*}(\bm x_i) &= \bm v_{y\rightarrow y^*}(\bm x_i,\bm \theta) + \bm\delta_v(\bm x_i) + \bm\epsilon_v(\bm x_i), ~~\bm\epsilon_v(\bm x_i) \sim \mathcal N(0,\sigma_v^2\bm I)\label{eqn:ve}.  
\end{align}
Notice the similarity of equation \ref{eqn:ye} and equation \ref{eq:fbmc}. Equation \ref{eqn:ye} is merely doing the functional response Bayesian model calibration of equation \ref{eq:fbmc} with aligned functions. Equation \ref{eqn:ve} is a standard multivariate response calibration of the shooting vectors. Doing one or the other of these calibrations will either calibrate to the shapes of curves or their timing, while using both ensures we match the shape and timing. This framework also allows us to separately specify the discrepancy in the aligned functional responses or their warping functions, which is a key benefit. For instance, if the discrepancy is a timing shift, that can be modeled directly through $\bm\delta_v(\bm x_i)$. A discrepancy that would add a shape characteristic to the curves could be modeled directly through $\delta_{\tilde y}(t,\bm x_i)$. Further, these two types of discrepancy could be used at the same time, making for a very practical discrepancy modeling framework.

In order to produce inferences with this model, we discretize $t$ so that equation \ref{eqn:ye} can be re-written in a vector form, $\tilde{ \bm z}(\bm x_i) = \tilde{\bm y}(\bm x_i, \bm\theta) + \bm\delta_{\tilde y}(\bm x_i) + \bm\epsilon_{\tilde z}(\bm x_i)$, and the joint likelihood can then be written as
\begin{align*}
    f \left( \tilde{\bm z}(\bm x_1),\dots,\tilde{\bm z}(\bm x_n), \bm v_{z\rightarrow y^*}(\bm x_1),\dots,\bm v_{z\rightarrow y^*}(\bm x_n)  ~|~    \bm\theta,\sigma^2_{\tilde z}, \sigma^2_v, \bm\beta_{\tilde y}, \bm\beta_v\right) = \\
    \prod_{i=1}^n 
    \left[
    \mathcal{N}\left(\tilde{\bm z}(\bm x_i) ~|~ \tilde {\bm y}(\bm x_i, \bm\theta) + \bm\delta_{\tilde y}(\bm x_i) ,~\sigma_{\tilde z}^2\bm I\right) ~
    \mathcal N\left(\bm v_{z\rightarrow y^*}(\bm x_i) ~|~ \bm v_{y\rightarrow y^*}(\bm x_i,\bm \theta) + \bm\delta_v(\bm x_i) ,~\sigma_v^2\bm I\right)
    \right]
\end{align*}
where $\bm\beta_{\tilde y}$ and $\bm\beta_v$ parameterize the discrepancy terms of $\tilde{y}$ and $v$ respectively.
With a prior specified for $\bm\theta$, $\sigma^2_{\tilde z}$, $\sigma^2_v$, $\bm\beta_{\tilde y}$, and $\bm\beta_v$, the posterior
$$\pi \left(\bm\theta,\sigma^2_{\tilde z}, \sigma^2_v, \bm\beta_{\tilde y}, \bm\beta_v ~|~ \tilde{\bm z}(\bm x_1),\dots,\tilde{\bm z}(\bm x_n), \bm v_{z\rightarrow y^*}(\bm x_1),\dots,\bm v_{z\rightarrow y^*}(\bm x_n) \right)$$ is proportional to likelihood multiplied by priors and can be sampled with MCMC. Note that this approach requires the warping decomposition (and computer model) to be called for each likelihood evaluation.  Even though this is just the decomposition of a single curve and fairly fast, this expense can make emulation more desirable.  

\subsubsection{Calibration when Emulation is Necessary}
Assume that the set of $N_{sim}$ model runs is decomposed into aligned functions and warping functions. We can then use separate or joint emulators with training inputs $\left\{\bm x_j, \bm u_j \right\}_{j=1}^{N_{sim}}$ and outputs $\left\{ \tilde{\bm y}(\bm x_j, \bm u_j), v_{y\rightarrow y^*}(\bm x_j,\bm u_j) \right\}_{j=1}^{N_{sim}}$.  A full Bayesian approach to emulation and calibration results in a joint likelihood of the observation data and model runs while a modular Bayesian approach fits the emulator first and then uses the emulator to perform the calibration.  Because these models result in likelihoods and posteriors that are not vastly different from the likelihood without an emulator, we omit the explicit likelihood here.  For Gaussian process emulators, the likelihood under the full Bayesian functional calibration model (in basis space) can be seen in \cite{higdon2008computer}. In fact, we have purposefully designed this model so that the inference could be carried out using practical frameworks like the ones in \cite{higdon2008computer,francom2019inferring,bayarri2007framework,gu2018scaled,williams2006combining}. Each of these approaches can be applied by merely replacing their use of the original functional responses $z$ and $y$ with aligned responses ($\tilde z$ and $\tilde y$) and shooting vectors ($v_{z\rightarrow y^*}$ and $v_{y\rightarrow y^*}$). That being said, there are a number of modeling choices and practical considerations that can make this approach successful that we discuss in the next section.

\begin{figure}[htbp]
\vspace{-.2in}
     \centering
     \begin{subfigure}[b]{\textwidth}
         \centering
         \includegraphics[width=\textwidth]{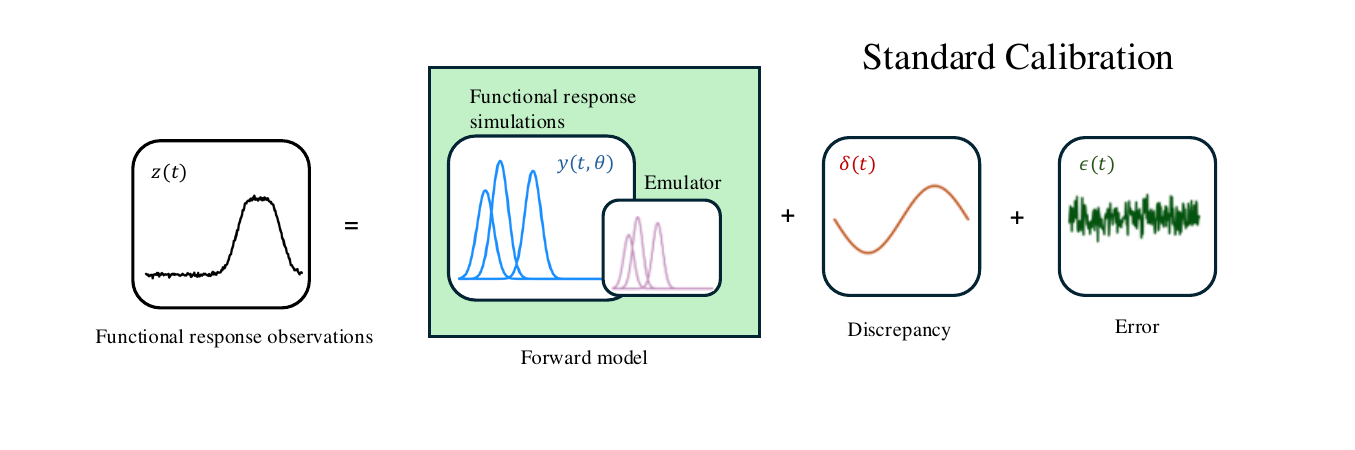}
         \vspace{-.5in}
         \caption{The standard approach to Bayesian model calibration with functional response models only amplitude variation with $z(t) = y(t,\theta)+\delta(t)+\epsilon(t)$.}
         \label{fig:diagram-standard}
     \end{subfigure}
     \hfill
     \begin{subfigure}[b]{\textwidth}
         \centering
         \vspace{.1in}
         \includegraphics[width=\textwidth]{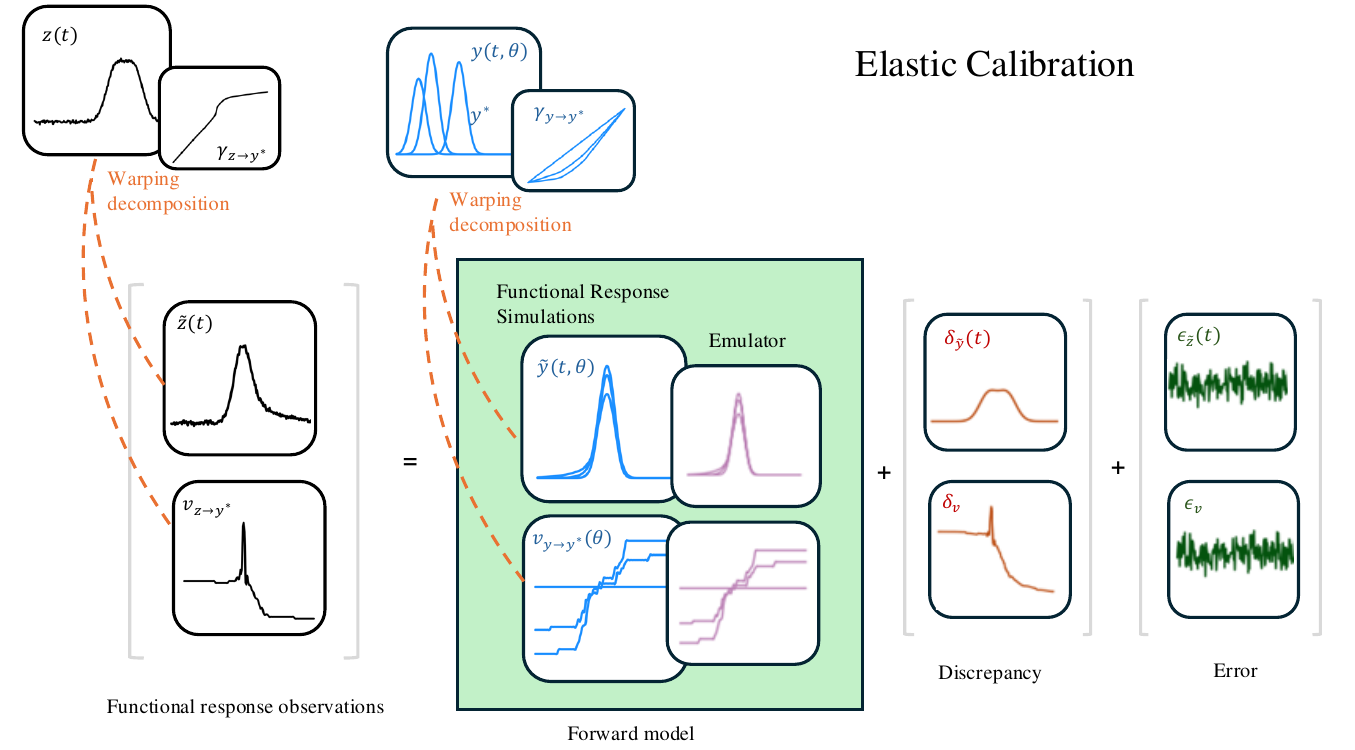}
         \caption{The elastic approach to Bayesian model calibration with functional response models amplitude variation with $\tilde z(t) = \tilde y(t,\bm\theta)+\delta_{\tilde y}(t)+\epsilon_{\tilde z}(t)$ and phase variation with $\bm v_{z\rightarrow y^*} = \bm v_{y\rightarrow y^*}(\bm\theta)+\bm\delta_{v}+\bm\epsilon_{v}$.}
         \label{fig:diagram-elastic}
     \end{subfigure}
     \vspace{-.3in}
      \caption{The standard approach to Bayesian model calibration with functional response (a) compared to the elastic approach (b). The elastic approach first isolates phase and amplitude variation as a preprocessing step, and then uses the same techniques used to do inference under the standard approach.}
        \label{fig:diagram}
\end{figure}

Figure \ref{fig:diagram} shows a schematic of what the different parts of the elastic Bayesian calibration model are (Figure \ref{fig:diagram-elastic}) compared to the standard approach to calibration with functional response (Figure \ref{fig:diagram-standard}). The warping decomposition is a preprocessing step that results in two datasets, after which standard functional response calibration can be applied using the two datasets. This is a simple approach, but we will demonstrate that it can result in calibration that is significantly more accurate.

\subsection{Modeling Choices and Practical Considerations} \label{sec:ebmc_choices}

To this point, we have discussed how using standard functional response calibration tools with misaligned functional responses is possible but prone to problems.  We also mentioned that emulation techniques for functional response do not perform well when applied to misaligned functional responses, as demonstrated in \cite{francom2022landmark}.  We then showed how elastic functional data analysis methods use proper distance metrics by aligning functions in SRVF space and using (1) distance between aligned functions and (2) distance between shooting vectors (transformed warping functions). Then we introduced how to frame functional response computer model calibration such that it uses these proper elastic metrics.  Simply put, we use the warping decomposition to separate our misaligned functional responses into aligned functional responses and shooting vectors, and we use these two new datasets instead of the original data when calibrating.

Below we discuss additional modeling choices and practical considerations for using this methodology.

\subsubsection{Warping decomposition} 

\noindent\textbf{\textit{Uncertainty:}}
Because we are using an optimization technique to obtain the warping functions, we are fixing a part of the model that could be considered uncertain.  This is a modeling choice, similar to the choice of \cite{kennedy2001bayesian} to fix some emulator parameters at maximum likelihood values or to the choice of \cite{higdon2008computer} to not allow for uncertainty in the basis representation of the functional response (i.e., the functional principal components are fixed). As with these other modeling choices, a potential consequence is that uncertainty is underestimated.  If the warping functions were to be inferred jointly with all of the other unknowns in a Bayesian framework, this would lead to a much greater computational burden and would require much more specialized calibration software. Similar to emulation, a possible shortcut could be to use modular Bayes techniques to propagate uncertainty from the warping decomposition to calibration uncertainty while cutting the feedback from the calibration to the warping decomposition \citep{liu2009modularization,plummer2015cuts}. 

\noindent\textbf{\textit{Regularization:}}
Another practical consideration is for the family of warping functions allowed in the warping decomposition, $\Gamma$.  If these functions are not smooth or regularized, their variation can be difficult to predict using the parameters, resulting in emulators with large residual variance.  However, if they are over regularized, they may not align the functional responses enough. 
In case one wants to control the amount of warping or {\it elasticity} this can be done 
as described in \cite{wu-srivastava:2011} using a penalty on equation~\ref{eq:optim}.
In the examples and applications below, we arrive at regularization choices by trial and error, where we are satisfied with a level of regularization when it results in nicely aligned functional responses with relatively smooth warping functions.

\noindent\textbf{\textit{Choice of alignment reference:}}
This decomposition can occur by either aligning the model to the experiment or by aligning to some other common element (e.g., one of the model runs). Alignment to the experiment is natural in many cases, but there are a few reasons why aligning to one of the model runs could be useful.  First, the model runs have no discrepancy or noise, which means that warping to them can be more stable.  Second, exploration of the computer model is often of interest even if there is no calibration data, in which case emulation is frequently used and can be performed more accurately when accounting for misalignment.

\noindent\textbf{\textit{Modeling transformed aligned curves:}}
Recall that we use the SRVF transformation of the original curves to obtain the warping decomposition.  However, we opt not to build the model for the aligned data in the SRVF space (i.e., we model $z$ and $y$ instead of $q_z$ and $q_y$).  We do this because modeling in derivative space means that when models are transformed back to the native space (e.g., to make predictions), errors are integrated and unrealistic heteroskedasticity arises.  Using $\ltwo$ distance between the aligned curves (rather than their SRVFs) still results in a proper metric, as long as the SRVFs were used for the alignment.
Note that we do not experience heteroskedasticity issues when transforming from shooting vectors to warping functions because of the highly-constrained shape of the warping functions.

\subsubsection{Emulation}
\cite{francom2022landmark} showed that taking alignment into account can improve emulator accuracy and efficiency, though their approach relied on a model for landmarks (rather than shooting vectors) to build warping functions.  They found that, in many cases, training separate emulators for aligned functions and warping functions is more desirable than training a single emulator for both.  This is because the warping function model will often have larger unexplained variation (i.e., larger residual variance than the aligned data model, even under appropriate standardization) due to the latent nature of the warping functions.  Subtle variations in the warping function regularization level can result in fairly significant variations in the shooting vectors, and while this variation is less pronounced when transforming back into warping function space, the larger variation can ``corrupt'' \citep{liu2009modularization} the modeling of the aligned data unless careful precautions are taken.  The easiest precaution is to merely create the two emulators independently.  This works well in practice.

\subsubsection{Discrepancy models}
The decomposition of misaligned data into aligned data and shooting vectors not only facilitates better emulation and calibration likelihood via proper distance metrics, it also facilitates more realistic discrepancy modeling.  For instance, if the discrepancy is a time shift, that can be expressed through the shooting vector discrepancy model $\bm \delta_v$.  If the discrepancy is a change to one feature of the curves, that can be handled directly through the aligned data discrepancy model $\bm\delta_{\tilde y}$.  The specification of $\bm\delta_{\tilde y}$ can be reasoned about in a natural way (compared to trying to specify discrepancy in the misaligned space), for instance a set of basis functions could be used as in \cite{higdon2008computer}.  However, the specification of $\bm \delta_v$ is more nuanced because it is in a transformed space.  For instance, adding a constant in that space (e.g., $\bm \delta_v(\bm x_i) = \bm 1$) has no effect because the shooting vectors are scaled by their norm in the exponential map that transforms them back to a SRVF on the unit sphere.  Hence, a constant time shift discrepancy is not achieved by adding a constant to the shooting vectors. In the simulated example with discrepancy in Section \ref{sec:sim_discrep}, we demonstrate modeling $\bm\delta_{\tilde y}$ and $\bm \delta_v$ with B-spline basis functions, and compare to the case where the discrepancy is only defined in the standard way.  In the flyer plane analysis below, we use a piecewise linear basis for $\bm \delta_v$ to capture time shifts.  In all basis expansions for $\bm \delta_v$ we exclude any intercept term, because constant shifts in shooting vectors are not useful.

\subsubsection{Residual error models}
We will typically assume independence between $\bm\epsilon_{\tilde z}$ and $\bm\epsilon_v$, but this can be relaxed. Of greater interest is the assumption of independence within $\bm\epsilon_{\tilde z}$ and $\bm\epsilon_v$.  We note here that adding a correlation structure to the residual model still results in appropriate distance metrics in the likelihood calculation.  This can be done explicitly to avoid the problems identified in \cite{brown2018}.

\subsection{Identifiability}

There are multiple places where identifiability is important to consider in Bayesian model calibration. In a full Bayesian approach, there are identifiabiliy concerns between estimating the parameters, discrepancy, emulator, and measurement error. Modularization can improve parts of this; for example, modularizing the emulator removes it as a contributor to the lack of identifiability \citep{liu2009modularization}. By separating amplitude and phase variability, allowing us to separately emulate them and separately include discrepancy and measurement error models, we improve identifiability. For example, if the discrepancy is a time shift (which is common), the only way to specify that in the standard calibration approach is to allow flexible discrepancy across amplitude space. Of course, that does not put the discrepancy model explicitly in the time domain, which will lead to a less constrained (and hence less identifiable) discrepancy in the amplitude domain. Our approach allows for specifying discrepancy limited to the time domain (or amplitude domain, or both), which improves identifiability. It should be noted that decomposing a curve into phase and amplitude variation is not necessarily identifiable (and is separate from all the sources of identifiability mentioned above). However, regularizing the warping functions makes the system identifiable. When used appropriately, any decrease in identifiability because of the warping transformation is easily offset by the better identified calibration model it enables.

\subsubsection{Using Standard Functional Calibration Tools for Elastic Calibration}
Our formulation results in a  Gaussian-likelihood pointwise-calibration using the aligned responses and the shooting vectors instead of the original misaligned functional responses. The likelihood function for the elastic calibration is formed by combining vector versions of equations \ref{eqn:ye} and \ref{eqn:ve}. This means that the same tools that were discussed in Section \ref{sec:pointwise_cal} can be used for elastic calibration, making the application of our proposed methods very practical.  For instance, SEPIA \citep{sepia} implements the model of \cite{higdon2008computer}, and \cite{impala} implements a similar model but allows for modularization of the emulator as in \cite{francom2019inferring} with a MCMC algorithm that handles posterior multimodality via tempering.  Both of these tools can be used to do elastic calibration, and the warping decomposition can be achieved using the {\it fdasrsf} R-package or equivalent python package \textit{fdasrsf} \citep{fdasrsf}.  

Further, there are several choices available to emulate computer model runs with functional response as described in \cite{francom2022landmark,hutchings2023comparing} and \cite{collins2024bayesian}. 
In the examples that follow, we use a Bayesian multivariate adaptive regression spline (BMARS) \citep{francom2018bass} emulator for both the aligned simulated curves and their corresponding shooting vectors, paired with the calibration approach in \cite{impala}. We chose BMARS due to its performance and the ability to scale to a large number of training points which we have in our equation of state and material strength data examples, though the methods are agnostic to emulator choice.

\section{Simulated Examples}\label{sec:simresults}

\subsection{Example With Comparison}
\begin{figure}[htbp]
	\centering
	\includegraphics[width=4in]{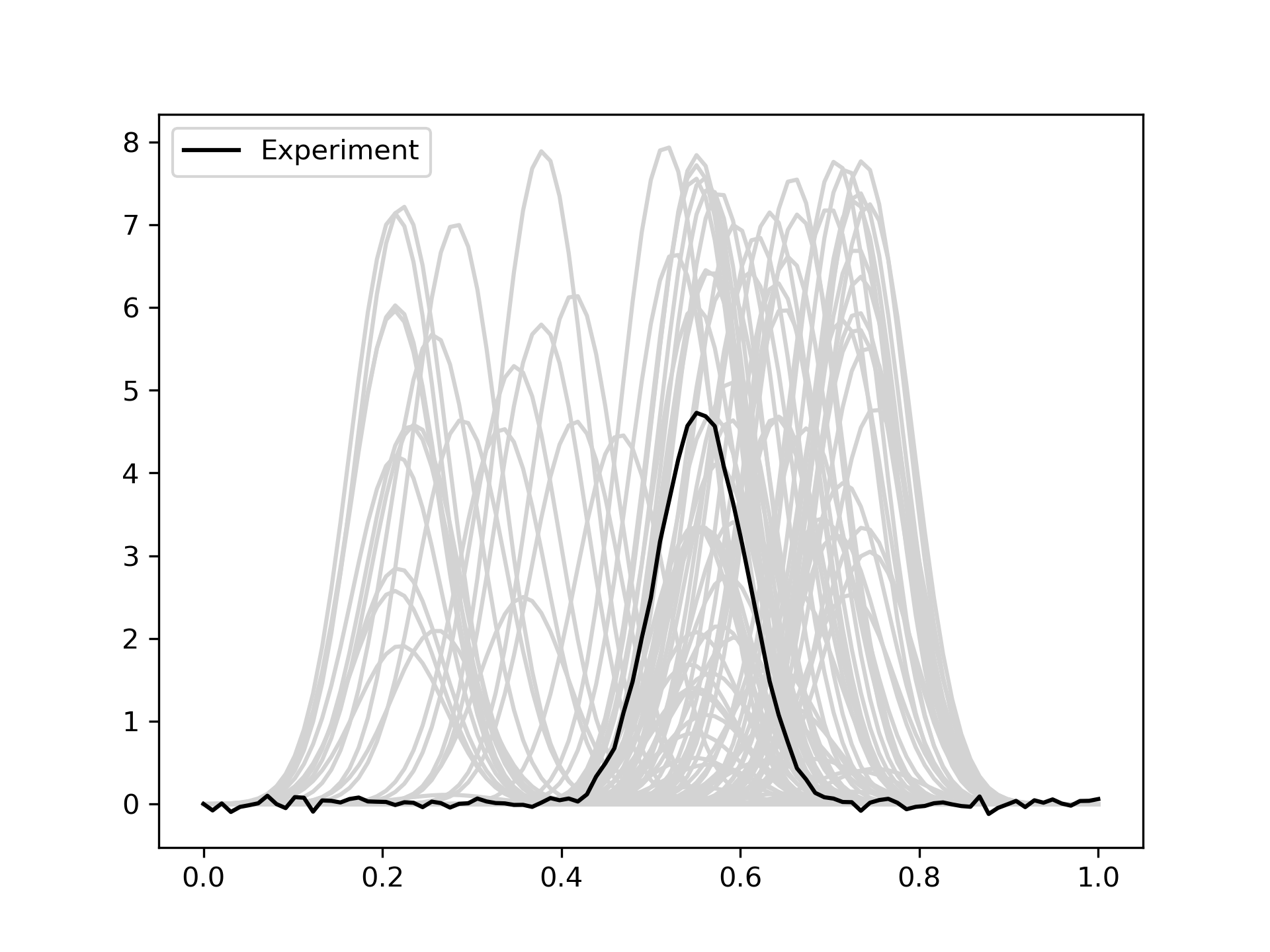}
	\caption{ Simulated curves $y(t,\bm u_1),\dots,y(t,\bm u_{100})$ and experiment data $z(t)$ for the simulated example.}
	\label{fig:sim_original}
\end{figure}

To illustrate the intuition behind this method, we simulate an example problem with three parameters to calibrate and misaligned functional response and no discrepancy (a similar example with discrepancy is included in the supplemental material).
The example is constructed using the following model to generate misaligned functional responses, where each function is a Gaussian pdf and the parameters control the location and height of the peak:
\[y(t,\bm u) = \frac{u_1}{0.05\sqrt{2\pi}}\exp\left(-\frac{1}{2}\left(\frac{t-(\sin(2\pi u_0^2)/4-u_0/10+0.5)}{0.05}\right)^2\right) + 0 u_2.\]
The calibration parameters are $\bm u=[u_0,u_1,u_2]$, though the functions only depend on the first two of these parameters. There are no experimental parameters so we omit $\bm x$ from the model specification used in the previous sections.
We generated a set of 100 functions (model runs) using $\bm u_1,\dots,\bm u_{100}$ where each $\bm u_j$ was sampled uniformly within the unit cube. The `experimental data' $z(t)$ was generated using the parameter values $\bm \theta=[0.1028, 0.5930, 0]$ and adding Gaussian noise with standard deviation 0.05, but without systematic bias (discrepancy).  Figure~\ref{fig:sim_original} presents the simulated model runs in grey and the experimental data shown in black.

\begin{figure}[htbp]
     \centering
     \begin{subfigure}[b]{0.49\textwidth}
         \centering
         \includegraphics[width=\textwidth]{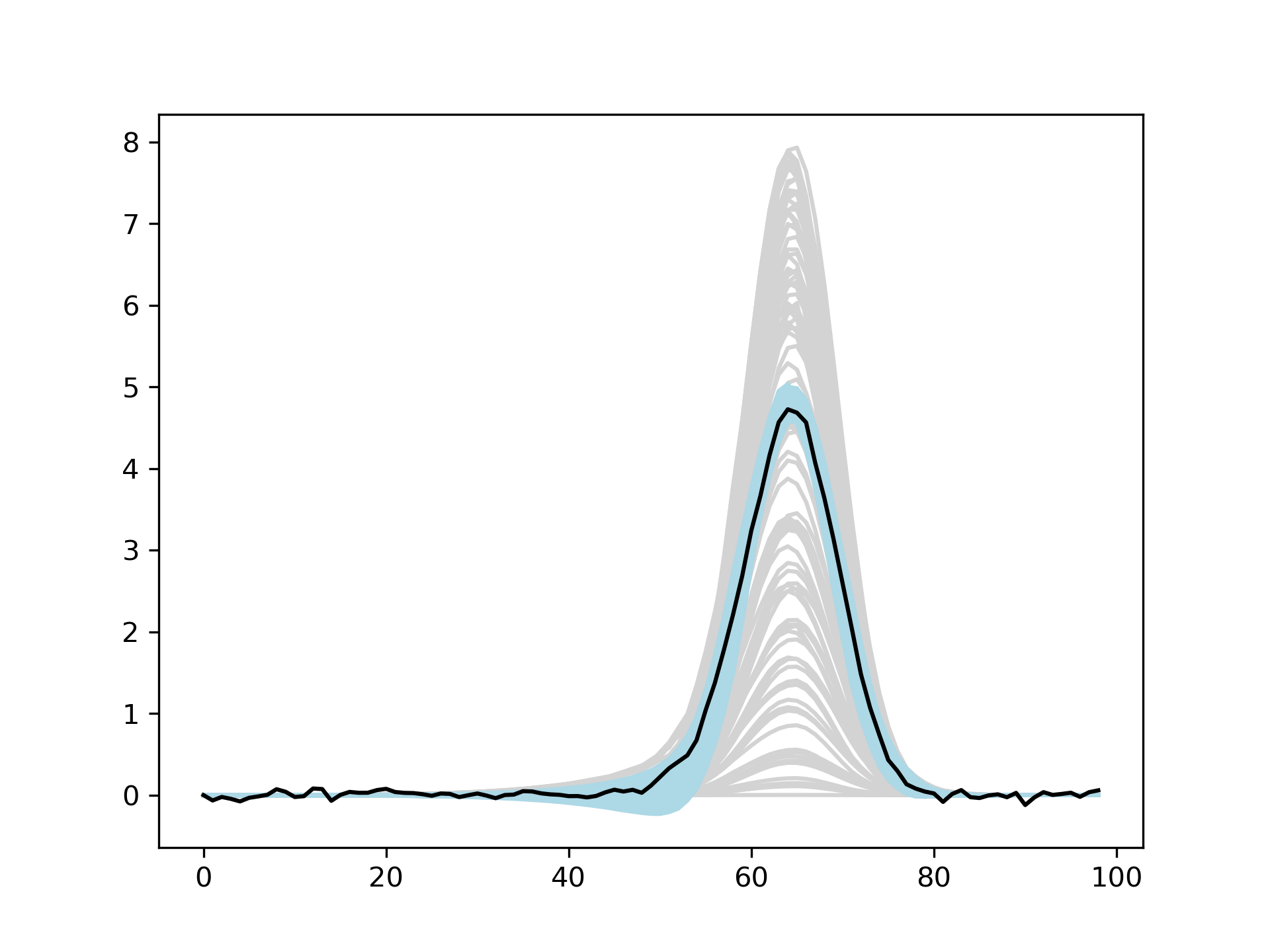}
         \caption{Calibrated aligned predictions.}
         \label{fig:sim_ftilde_cal}
     \end{subfigure}
     \hfill
     \begin{subfigure}[b]{0.49\textwidth}
         \centering
         \includegraphics[width=\textwidth]{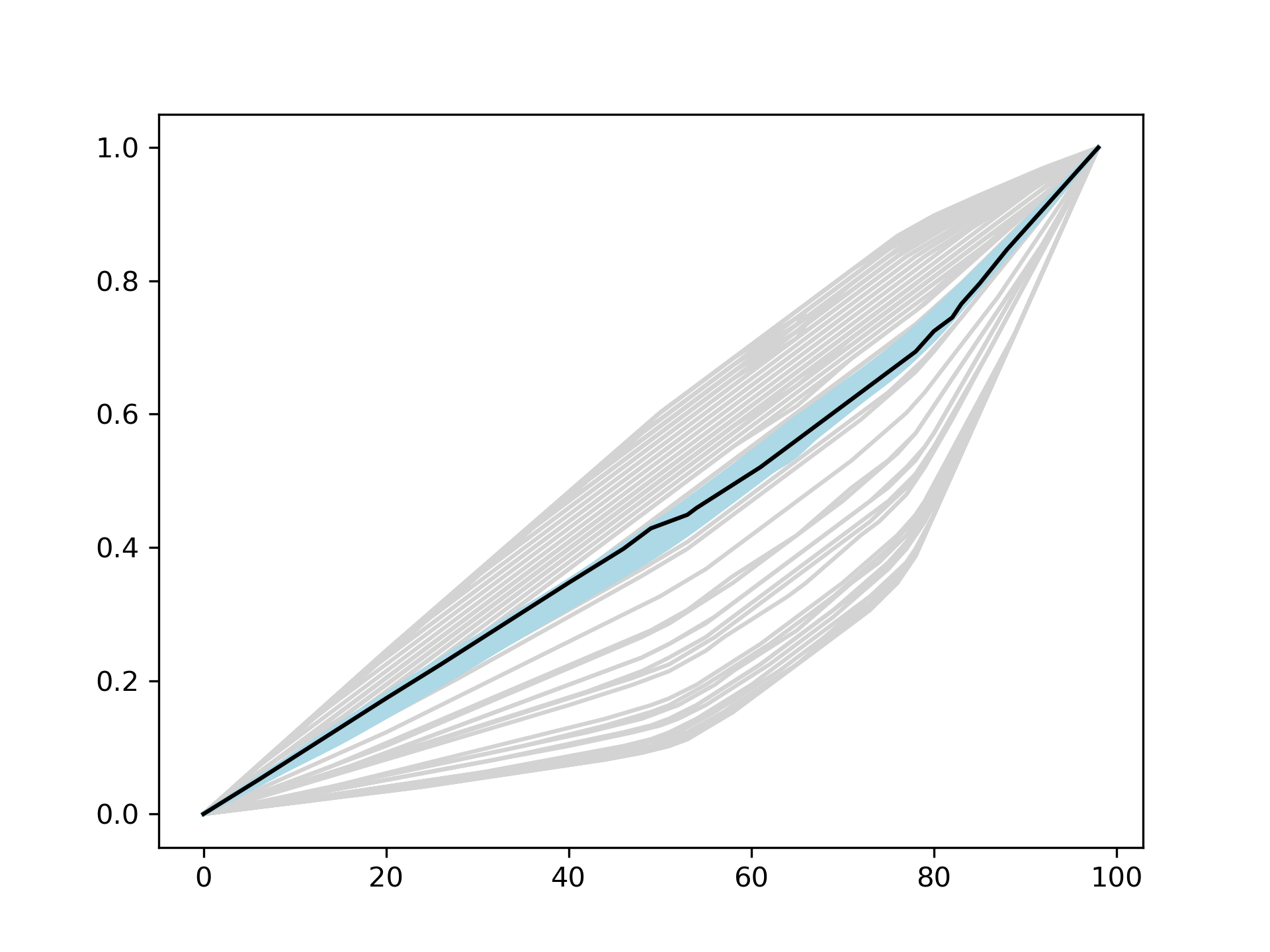}
         \caption{Calibrated warping functions.}
         \label{fig:sim_gam_cal}
     \end{subfigure}
     \begin{subfigure}[b]{0.49\textwidth}
         \centering
         \includegraphics[width=\textwidth]{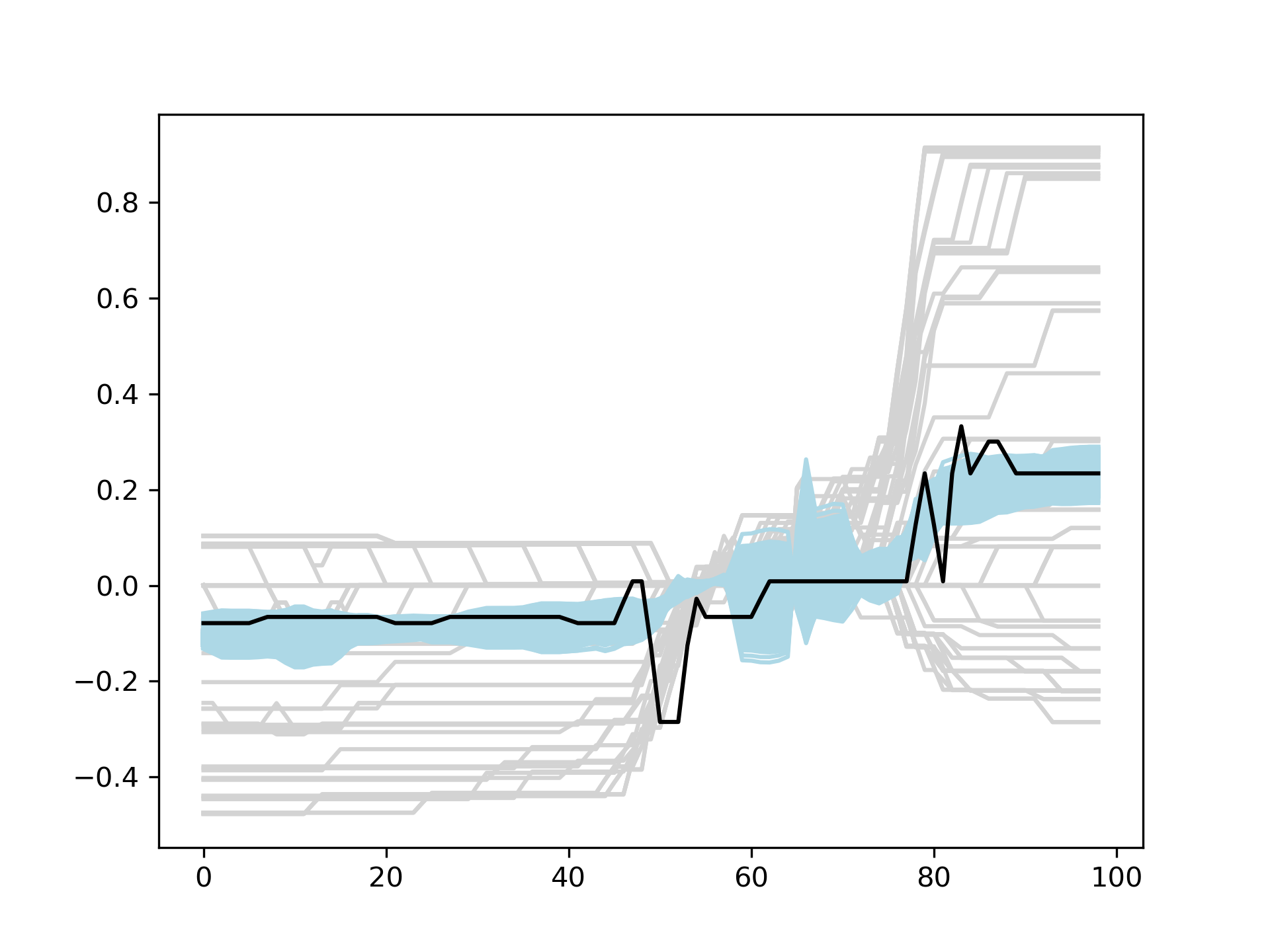}
         \caption{Calibrated shooting vectors.}
         \label{fig:sim_v_cal}
     \end{subfigure}
           \begin{subfigure}[b]{0.49\textwidth}
         \centering
         \includegraphics[width=\textwidth]{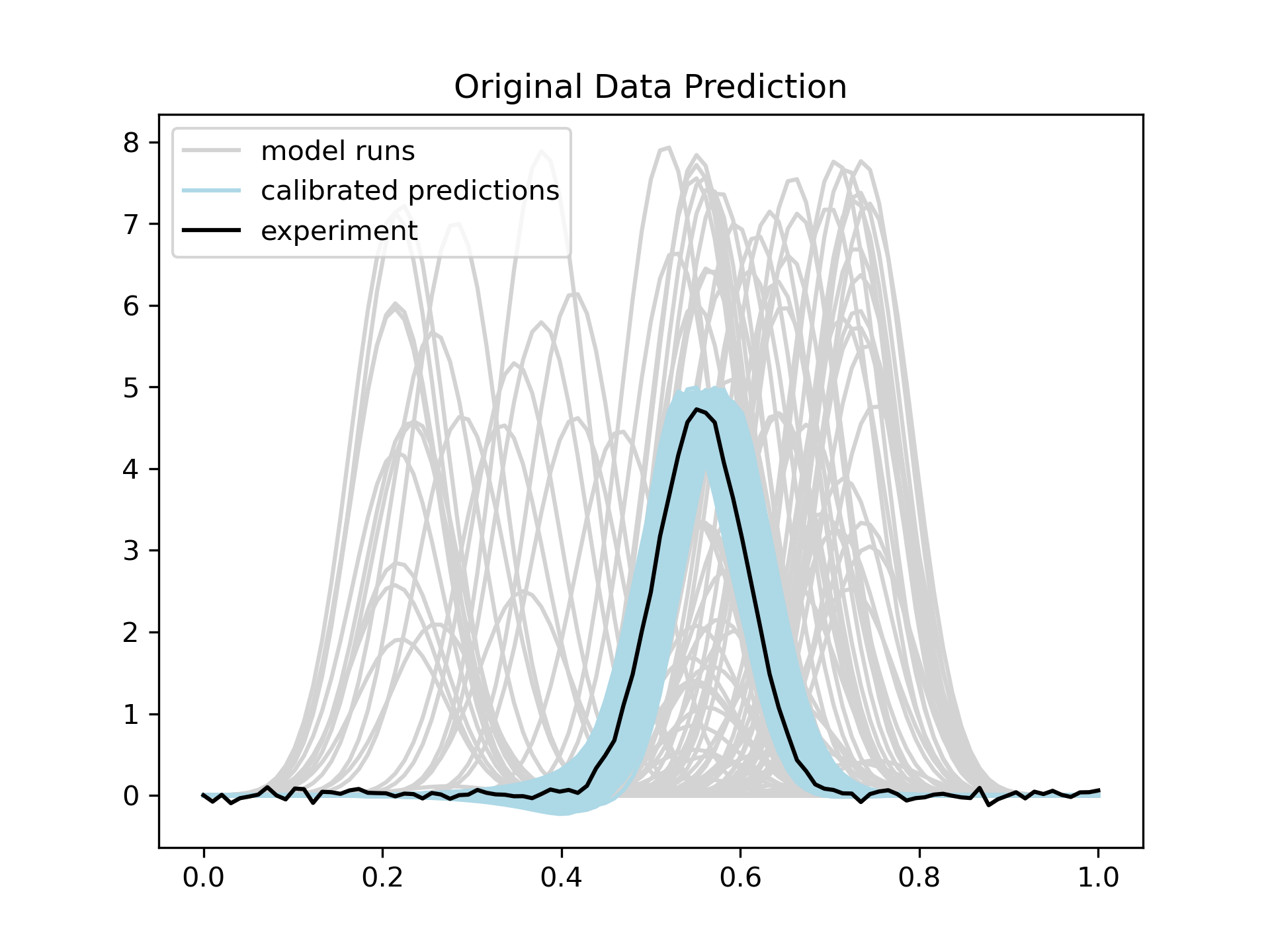}
         \caption{Calibrated misaligned predictions.}
         \label{fig:sim_calibration}
     \end{subfigure}
        \caption{Posterior predictive samples after calibration of the simulated data for Example 1.}
        \label{fig:sim_calibration_samples}
\end{figure}

We separated the phase and amplitude by aligning the simulated curves to one of the model runs utilizing the elastic methodology from Section~\ref{sec:align}. 
Figure~\ref{fig:sim_ftilde_cal} shows the aligned model runs (grey) and the experimental data (black), while Figure~\ref{fig:sim_gam_cal} presents the corresponding warping functions of both the simulated curves ($\gamma_{y\rightarrow z}(t,\bm u_j)$ for $j=1,\dots,100$) and the experimental data ($\gamma_{z\rightarrow z}(t)=t$). Figure~\ref{fig:sim_v_cal} presents the corresponding shooting vectors for the simulation ($\bm v_{y\rightarrow z}(\bm u_j)$) and the experimental data ($\bm v_{z\rightarrow z}$).

We built separate BMARS emulators for the aligned curves and the shooting vectors and performed a (modular) elastic Bayesian model calibration as described in Section~\ref{sec:ebcalibration} to infer the posterior distribution of $\bm \theta$ along with posterior predictions of the curves.

\begin{figure}[htbp]
	\centering
	\includegraphics[width=4in]{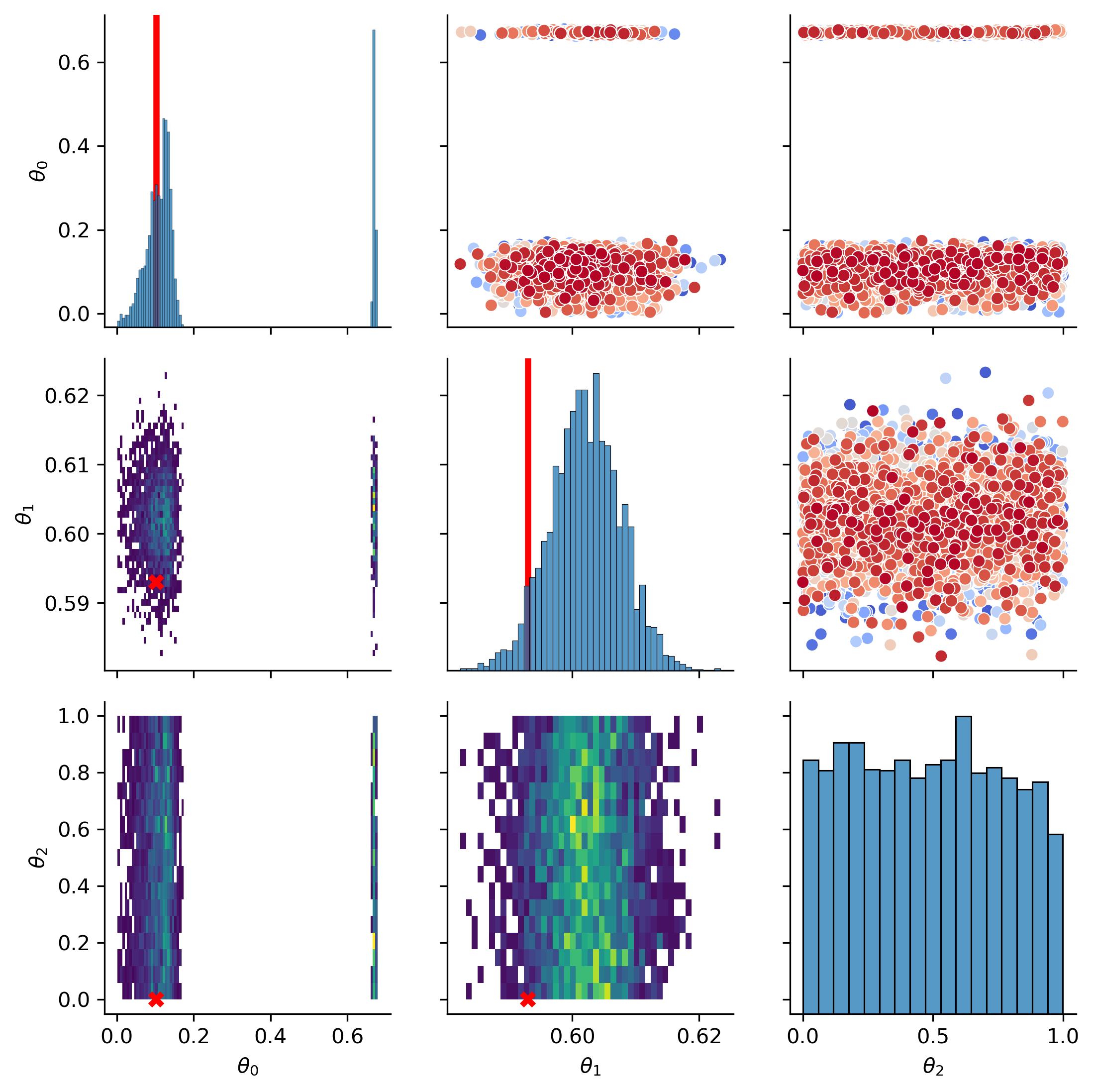}
	\caption{Pairs plot of parameters after calibration to simulated example data with the 'true' parameter combination shown by the x and vertical lines}
	\label{fig:sim_pairplot}
\end{figure}

Figure~\ref{fig:sim_calibration_samples} presents some posterior predictive samples after calibration of the aligned curves, the corresponding warping functions and shooting vectors (in blue), with their simulations (in grey) and for the experimental data (in black). Similarly, Figure~\ref{fig:sim_calibration} presents some posterior predictive samples in the original data space (in blue) with the simulated curves/model runs (in grey) and the experimental data (in black). The predictive samples cover the experimental data with small uncertainty, showing good predictive performance.
A pairwise plot summarizing the posterior samples of $\bm \theta$ (marginal and bivariate) is shown in Figure~\ref{fig:sim_pairplot}. 
The true value of $\bm \theta$  is represented by the  mark (x) in the lower triangular panels and the vertical line on the diagonal panels.  The true parameter values are covered by high density regions of the posterior distribution, indicating they are recovered well by the model calibration.  Additionally, the posterior distribution of the nuisance parameter $\theta_2$ resembles a uniform distribution (prior) and does not affect the calibration. We note that the construction of the test function results in bimodal posterior for $\theta_0$.

\begin{figure}[htbp]
	\centering
	\includegraphics[width=4in]{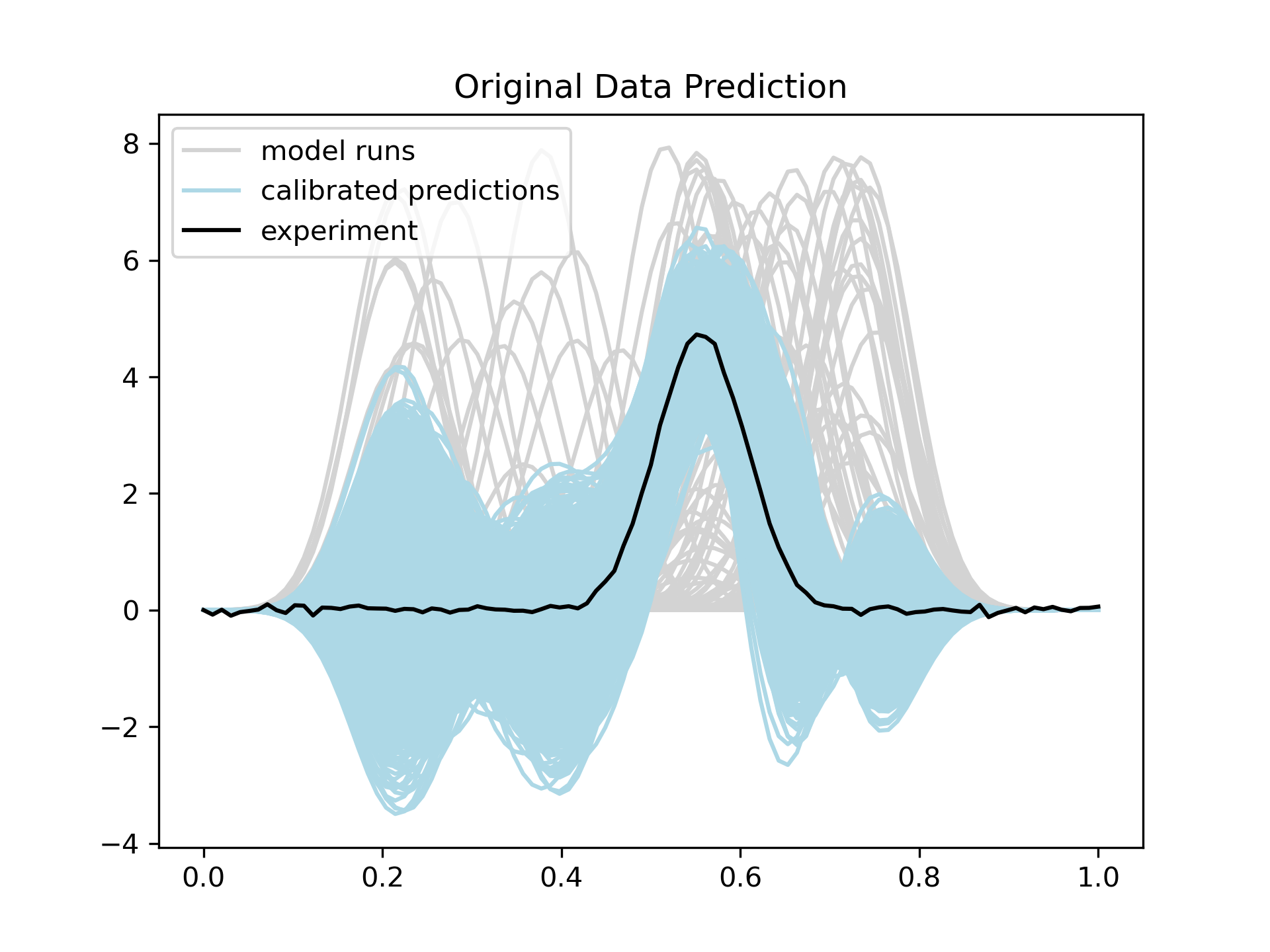}
	\caption{Calibrated predictions under the Bayesian calibration that does not account for misalignment.}
	\label{fig:sim_no_align_calibration}
\end{figure}

To compare our elastic Bayesian calibration method to a standard functional Bayesian calibration approach, we fitted a BMARS emulator on the original simulated curves and directly performed a modular Bayesian model calibration with no alignment. 
Figure~\ref{fig:sim_no_align_calibration} presents the posterior predictive samples after calibration of the simulated data in the original data space. 
We observe that the predictive samples do not resemble the experimental data well and have more than one mode/valley and large uncertainty, given that phase variability is not taken into account. The primary reason for the improved performance of the elastic calibration in this simulation is the improved emulator accuracy. The elastic emulator root mean squared error (RMSE) on test data is 0.082 versus the standard emulator RMSE of 0.562 on the same test data.

\begin{figure}[htbp]
	\centering
	\includegraphics[width=.5\textwidth]{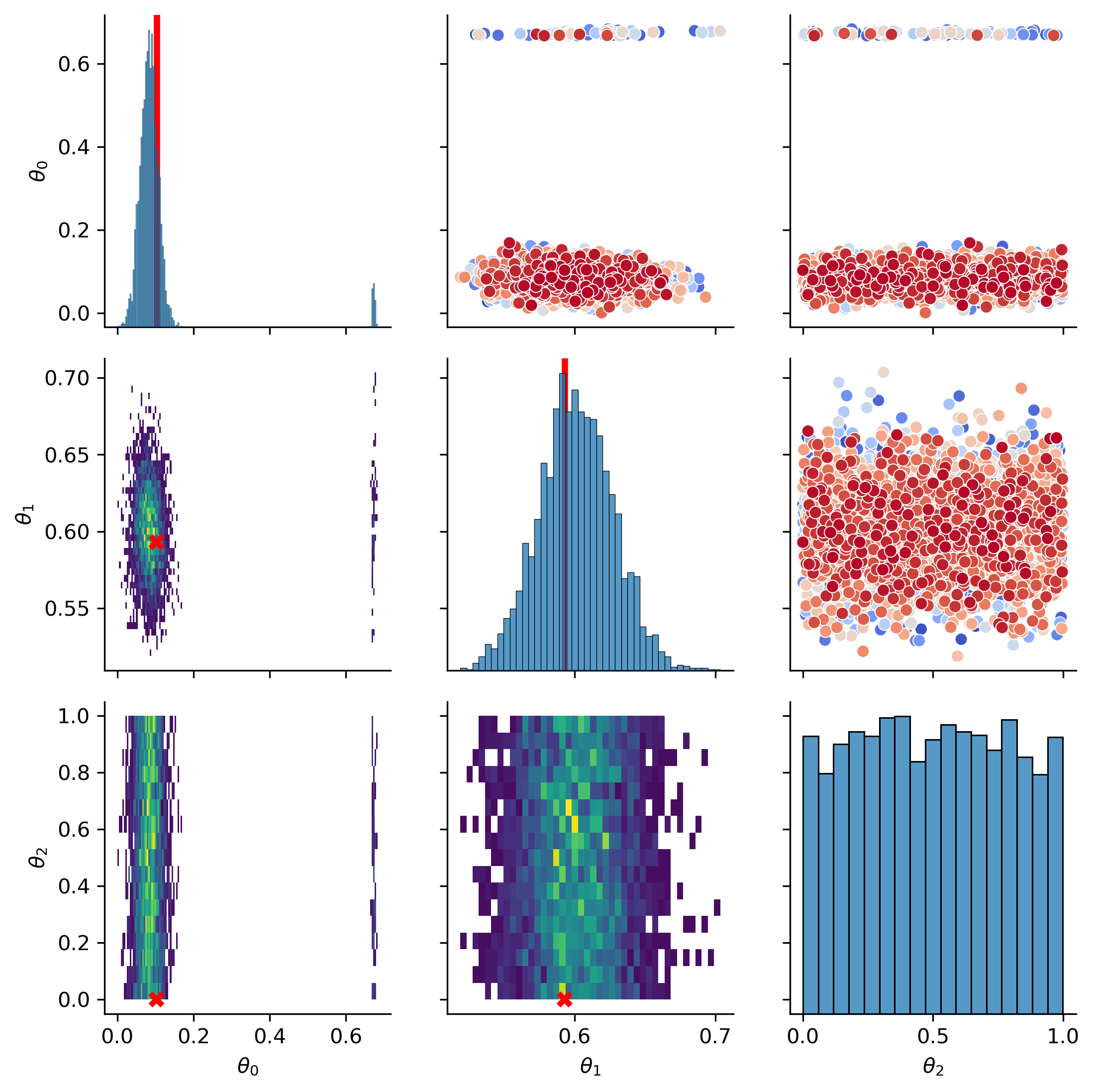}
	\caption{Pairs plot of the posterior samples of the parameters after standard calibration for the simulated example where the true value is marked by the  x and vertical line. The emulator was fitted directly on the original data. Note the axis limits are different from  Figure \ref{fig:sim_pairplot}.}
	\label{fig:sim_noalign_pairplot}
 \end{figure}

\begin{figure}[htbp]
	\centering
	\includegraphics[width=6in]{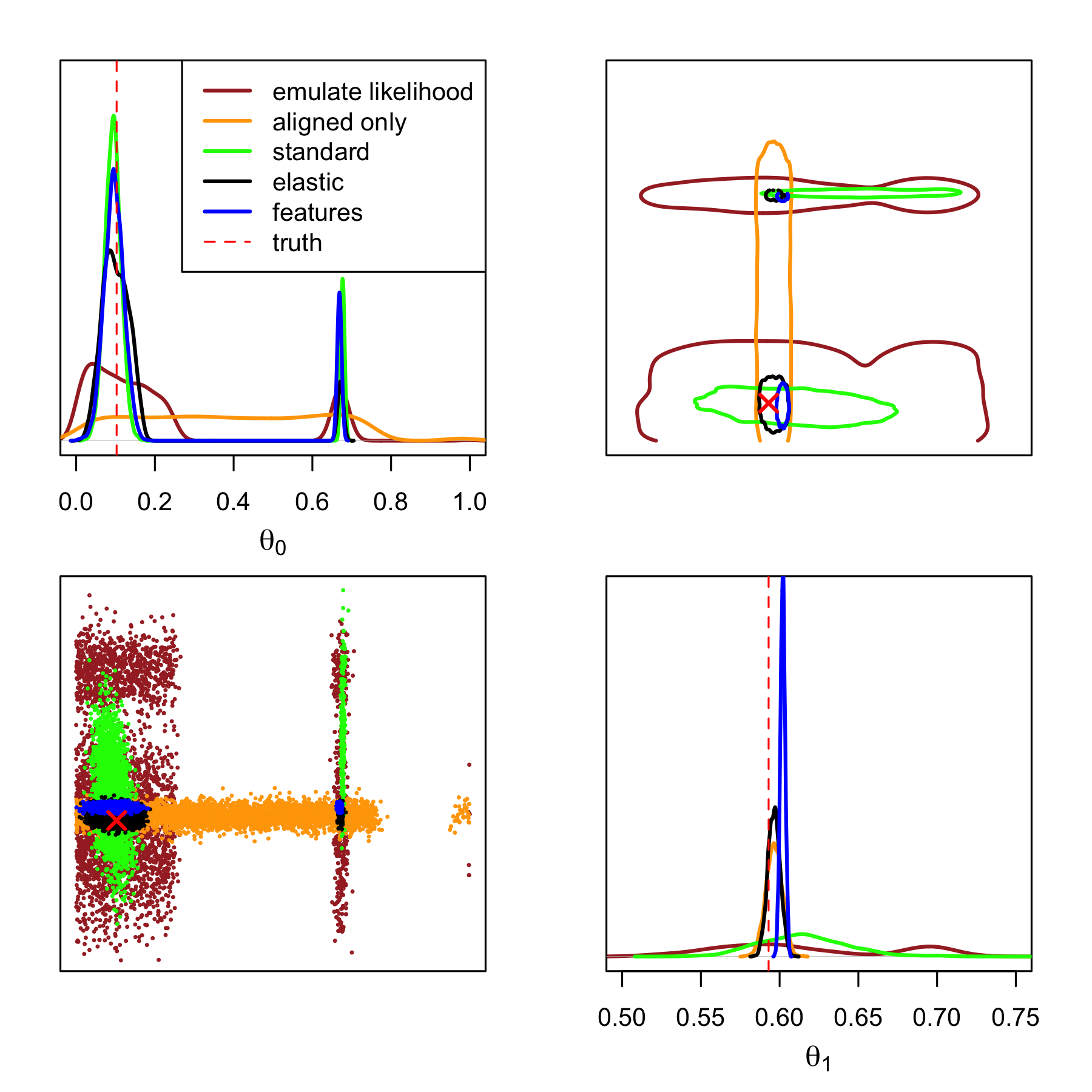}
	\caption{Comparison of the elastic calibration posterior samples with the posterior samples from four other approaches on the simulated example.}
	\label{fig:compare4}
\end{figure}

Beyond comparing to the standard functional Bayesian calibration approach, we also compare to calibration using features, calibration only using the aligned data, and calibration where the emulator is for the likelihood instead of the forward model. The results of these approaches are compared in Figure~\ref{fig:compare4}, which shows the samples of $\theta_0$ and $\theta_1$ under each method (the samples of $\theta_2$ are uniform in each case and are omitted from the figure) in the lower left. The true setting of $\theta_0$ and $\theta_1$ is denoted with a red X.  The top left and bottom right plots shows the marginal posterior density for $\theta_0$ and $\theta_1$ respectively. The top right shows the 95\% contour of the joint samples. There are many insights from this figure. When emulating the likelihood, we evaluated the mean squared error between the training sample aligned curves and the experiment aligned curve, repeated the same for the shooting vectors, built a joint BMARS emulator for the two mean squared errors, and calibrated to the zero vector. There are various ways to explore this further, but they are not the focus of this paper. This emulator fits well, but clearly not well enough to capture all of the information that the elastic approach does. The calibration that only uses aligned data is good at capturing $\theta_1$, but loses most information about $\theta_0$. The standard and elastic approaches are those shown in Figures \ref{fig:sim_noalign_pairplot} and \ref{fig:sim_pairplot}, respectively, but put on the same axes we see that the standard approach loses much information about $\theta_1$. For the feature-based approach, we selected two features: the maximum of each curve and the timing of the maximum. Given the form of the functions, this should communicate all of the information about $\theta_0$ and $\theta_1$. However, because the observations were noisy, the features for the observed curve were slightly biased, and this biased the calibration. In practice, features are usually not easy to select, are frequently noisy, and usually do not capture all of the information in the model, so that the feature-based calibration could be biased and lose information.  To summarize this comparison, all five approaches started with the same data, but the elastic approach resulted in the smallest uncertainty that still captured the truth. 

\subsection{Example With Discrepancy} \label{sec:sim_discrep}
\begin{figure}[htbp]
	\centering
	\includegraphics[width=4in]{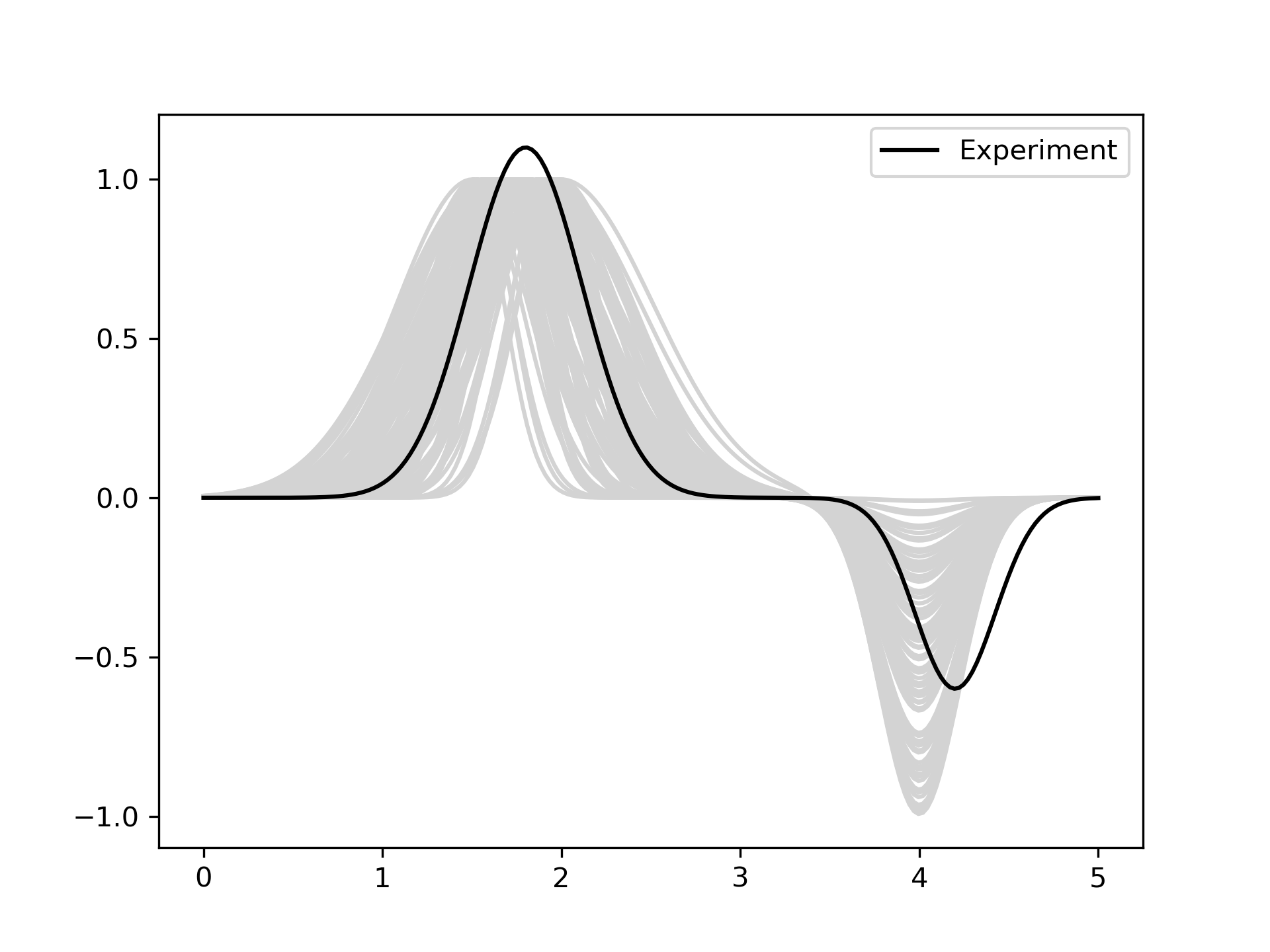}
	\caption{ Simulated curves $y(t,\bm u_1),\dots,y(t,\bm u_{200})$ and experiment data $z(t)$ for the second simulated example.}
	\label{fig:sim_original2}
\end{figure}

\begin{figure}[htbp]
	\centering
	\includegraphics[width=6in]{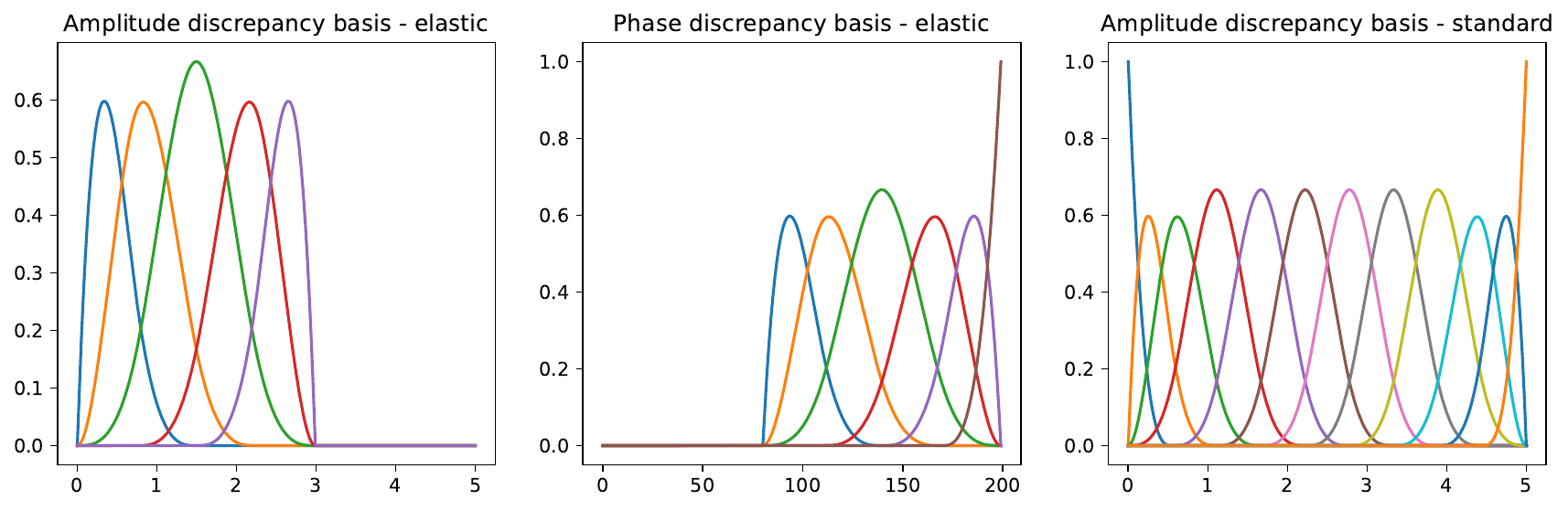}
	\caption{Basis functions defining $\bm D_{\tilde y}$ (left), $\bm D_{v}$ (middle), and basis functions used for standard calibration (right).}
	\label{fig:sim_D_basis}
\end{figure}

\begin{figure}[htbp]
	\centering
	\includegraphics[width=5in]{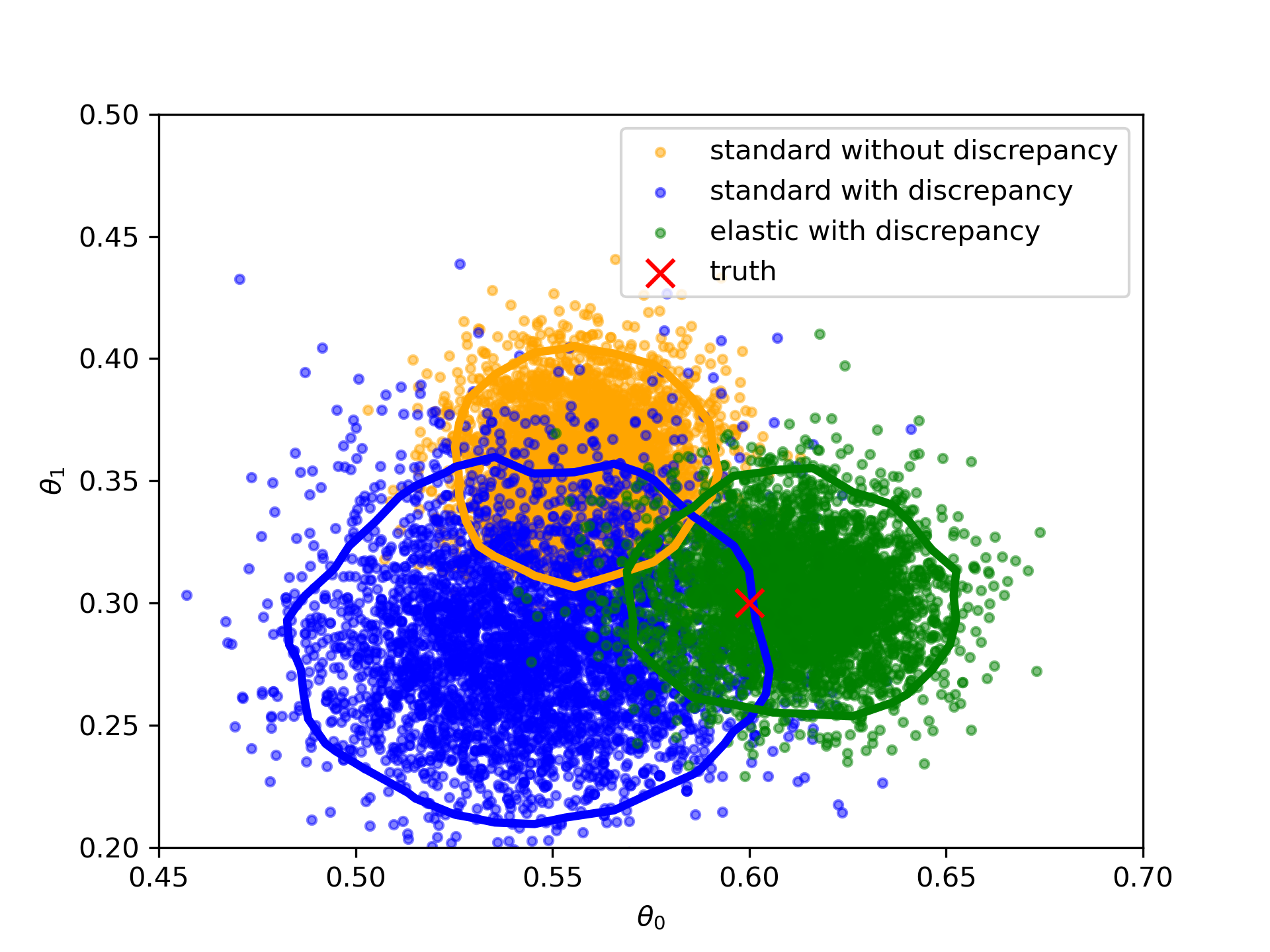}
	\caption{Comparison of elastic and standard inference of the calibration parameters. The points are posterior samples, while the lines are the 95\% contour for each posterior. Properly including the phase discrepancy in the elastic model allows us to get more accurate calibration than the standard approach which only includes amplitude discrepancy.}
	\label{fig:sim_theta_scatter3}
\end{figure}

\begin{figure}[htbp]
     \centering
     \begin{subfigure}[b]{0.49\textwidth}
         \centering
         \includegraphics[width=\textwidth]{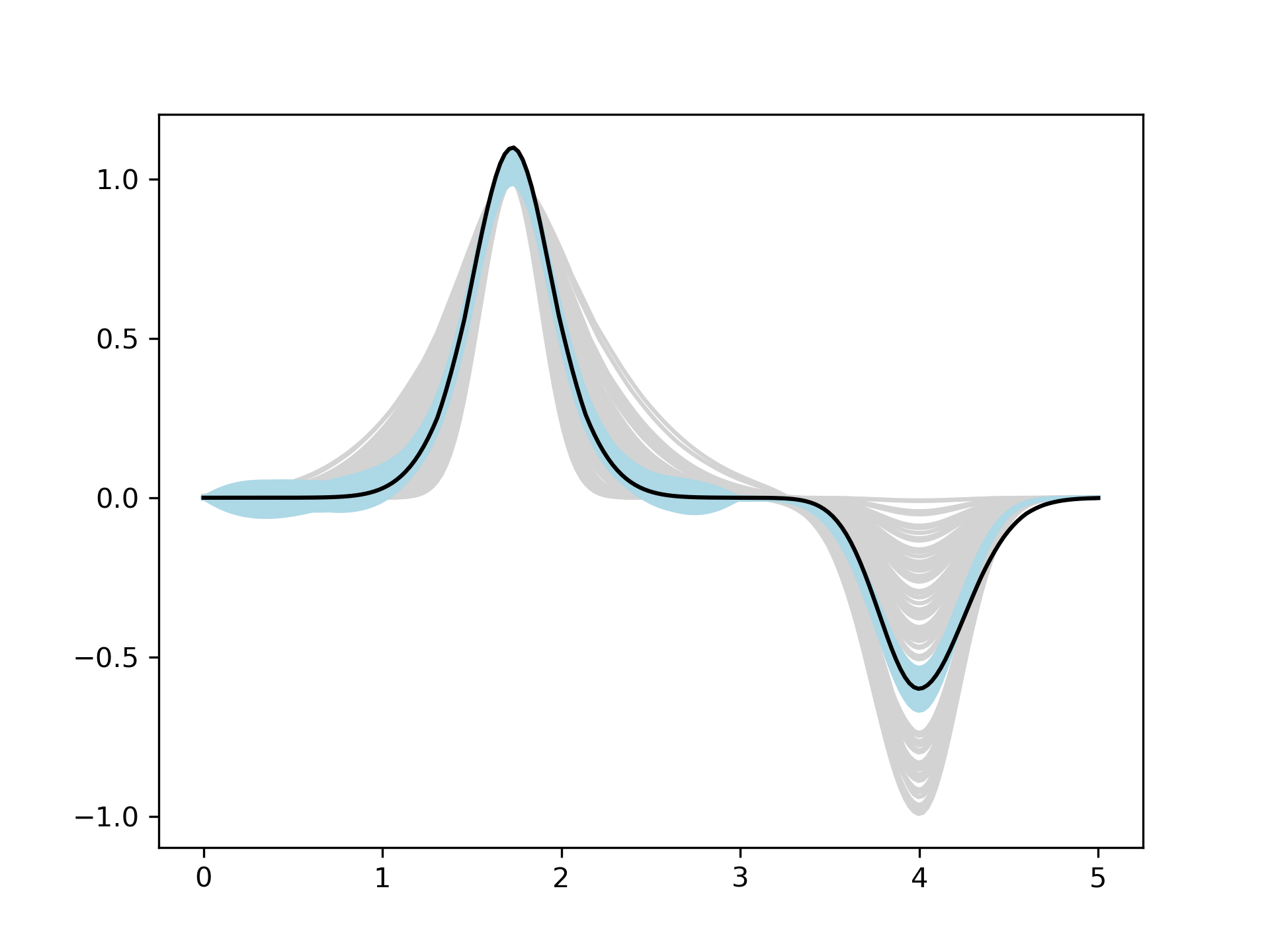}
         \caption{Calibrated aligned curves.}
         \label{fig:sim_ftilde_cal3}
     \end{subfigure}
     \hfill
     \begin{subfigure}[b]{0.49\textwidth}
         \centering
         \includegraphics[width=\textwidth]{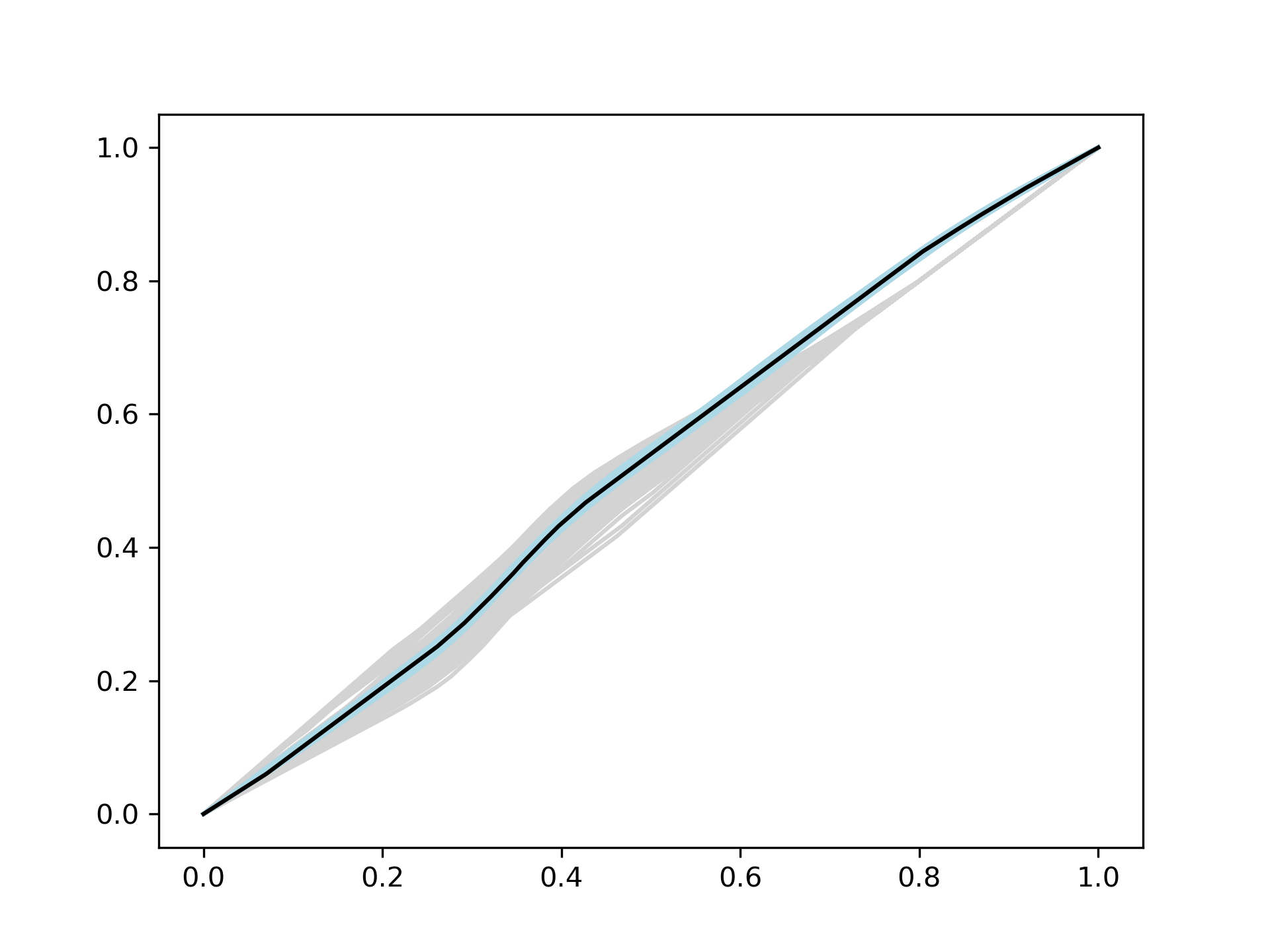}
         \caption{Calibrated warping functions.}
         \label{fig:sim_gam_cal3}
     \end{subfigure}
     \begin{subfigure}[b]{0.49\textwidth}
         \centering
         \includegraphics[width=\textwidth]{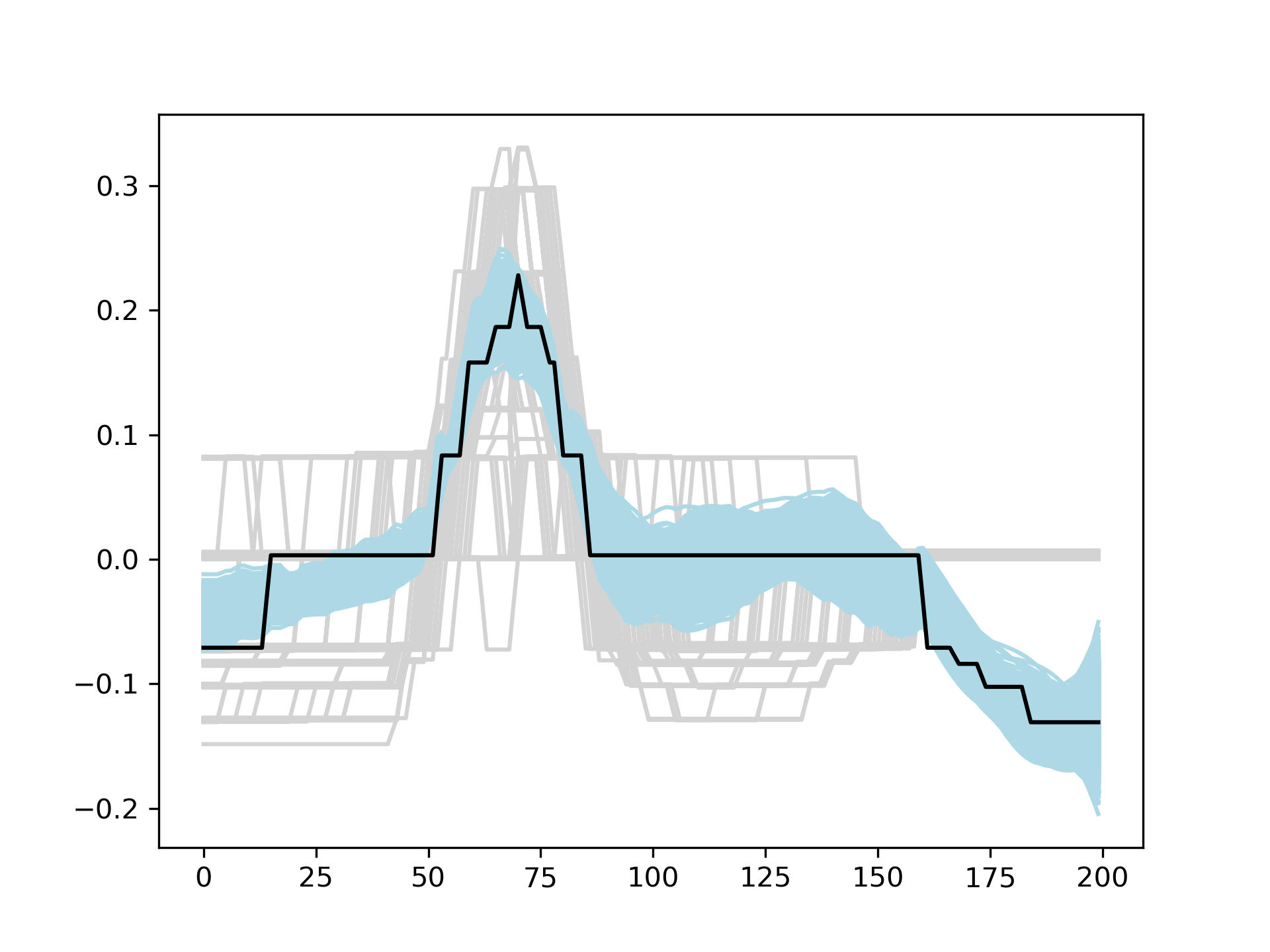}
         \caption{Calibrated shooting vectors.}
         \label{fig:sim_v_cal3}
     \end{subfigure}
           \begin{subfigure}[b]{0.49\textwidth}
         \centering
         \includegraphics[width=\textwidth]{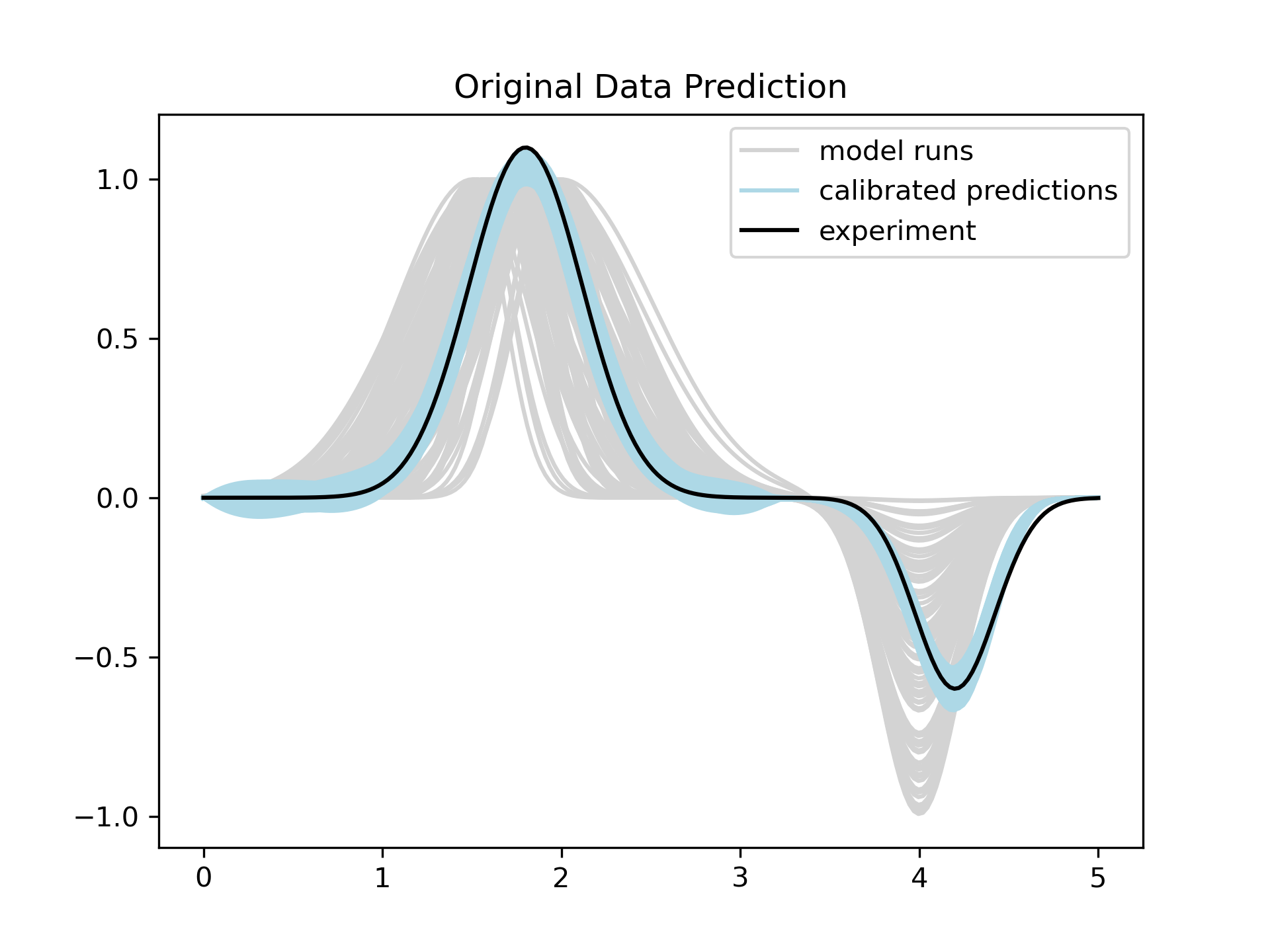}
         \caption{Calibrated misaligned predictions.}
         \label{fig:sim_calibration_discrep3}
     \end{subfigure}
        \caption{Posterior predictive samples after calibration of the simulated data for Example 2.}
        \label{fig:sim_calibration_samples3}
\end{figure}

\begin{figure}[htbp]
     \centering
     \begin{subfigure}[b]{0.49\textwidth}
         \centering
         \includegraphics[width=\textwidth]{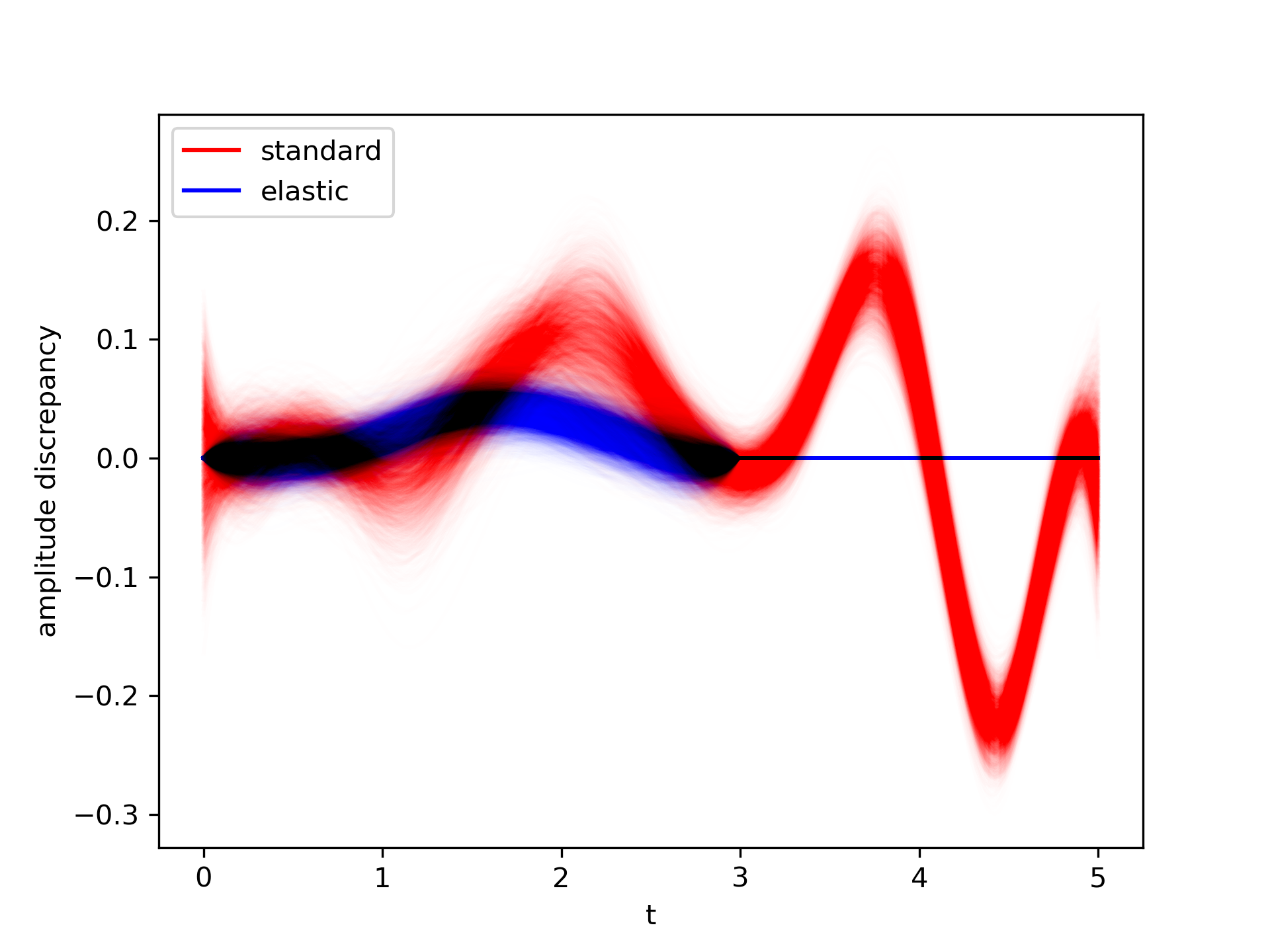}
         \caption{Amplitude discrepancy under the standard model and the elastic model.}
         \label{fig:amp_discrep}
     \end{subfigure}
     \hfill
     \begin{subfigure}[b]{0.49\textwidth}
         \centering
         \includegraphics[width=\textwidth]{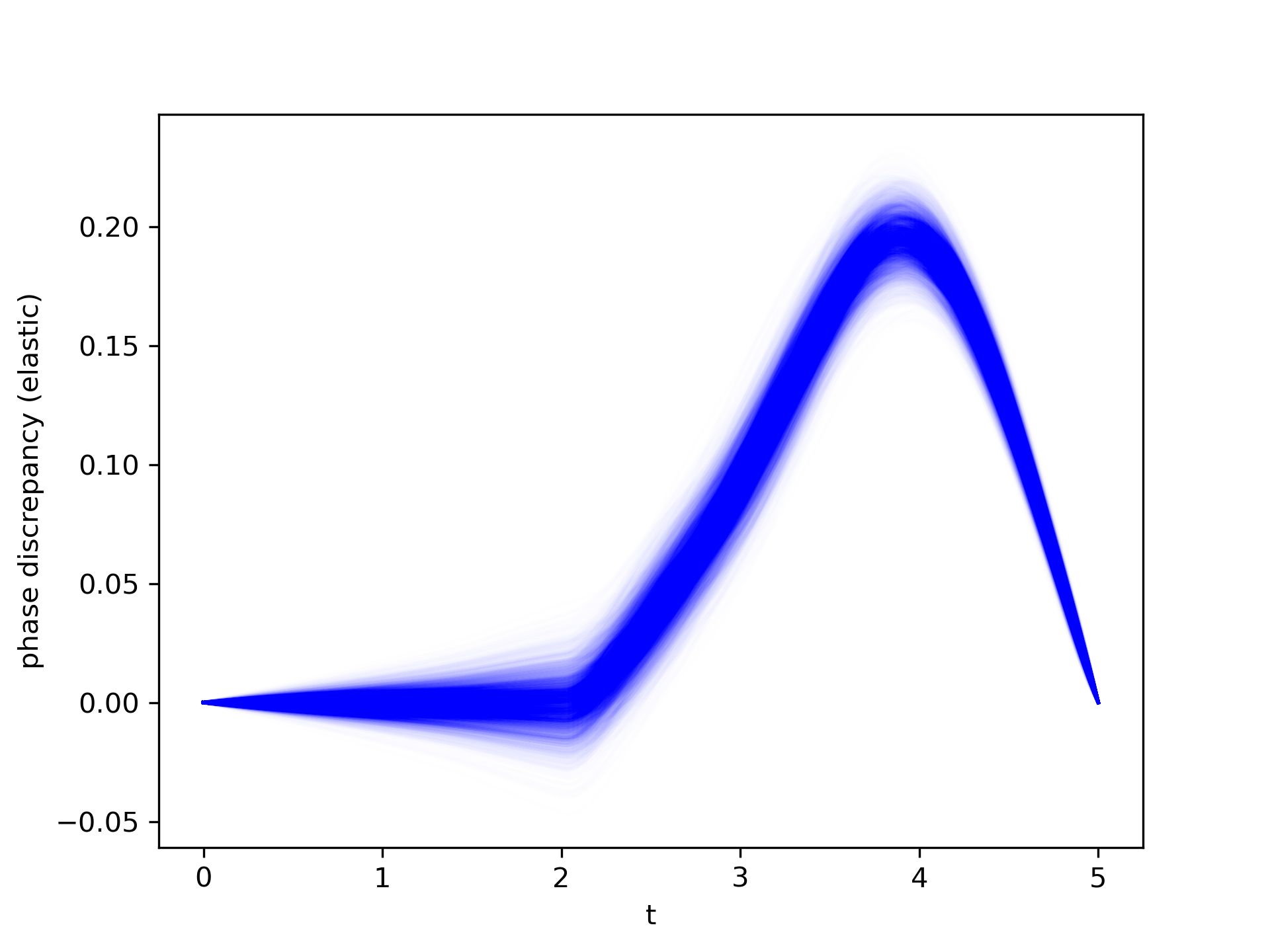}
         \caption{Phase discrepancy using elastic calibration.}
         \label{fig:phase_discrep}
     \end{subfigure}
      \caption{Comparison of standard and elastic discrepancy modeling for simulated example.}
        \label{fig:sim_discrep}
\end{figure}

In this example, we demonstrate the kind of discrepancy modeling that can be included in elastic Bayesian model calibration. We use the function
\[y(t,\bm u) = \exp\left\lbrace -2(t-0.5u_0-1.5)^2/(u_1+0.1) \right\rbrace - u_0\exp\left\lbrace -10(t-4)^2  \right\rbrace \]
which consists of two bumps, a positive bump with shifted location and a negative bump with shifted scale. There is no $x$ variable here for simiplicity, though the examples in the next section include $x$ variables. We generate 200 model runs using values of $\bm u$ from the unit square. We treat as our true function
\[z(t) = 1.1\exp\left\lbrace -2(t-0.5(0.6)-1.5)^2/(0.3+0.1) \right\rbrace - 0.6\exp\left\lbrace -10(t-4.2)^2  \right\rbrace \]
which is the simulator evaluated at $u_0=0.6,~u_1=0.3$ with two forms of discrepancy: the true function has larger first bump and shifted second bump. We include no measurement error. The model runs and true function are shown in Figure \ref{fig:sim_original2}. The discrepancy is such that the parameters cannot replicate the behavior of the true function. The challenge of discrepancy modeling will be to do it in such a way that our parameter inference is not biased. Because of the timing discrepancy in the second bump, traditional functional response calibration will struggle because an amplitude-only discrepancy will essentially lead to us losing information from the second bump. We use basis expansions for each discrepancy model so that $\bm \delta_{\tilde y} = \bm D_{\tilde y} \bm \beta_{\tilde y}$ and $\bm \delta_{v} = \bm D_{v} \bm \beta_{v}$. If our discrepancy was indexed by variable $\bm x$ in this problem, we would use $\beta_{\tilde y}(\bm x)$ and $\beta_{v}(\bm x)$ in place of $\bm\beta_{\tilde y}$ and $\bm\beta_{v}$. For both $\bm D_{\tilde y}$ and $\bm D_{v}$ we use B-spline basis functions shown in Figure \ref{fig:sim_D_basis}. The first panel in Figure \ref{fig:sim_D_basis} shows that we limit our discrepancy basis functions to the part of $t$ that corresponds to the first bump, because we know that is where the amplitude discrepancy is. The second panel in Figure \ref{fig:sim_D_basis} shows that we limit our phase discrepancy to the timing of the second bump because we know there is phase discrepancy there. When we perform calibration with these discrepancy models, we are able to recover the settings of $\bm\theta$ that generated the data, as shown in Figure \ref{fig:sim_theta_scatter3}, while standard calibration is more likely to get biased parameter inference. Figure \ref{fig:sim_calibration_samples3} shows the posterior predictions of our aligned data, shooting vectors, warping functions, and misaligned data when discrepancy are included and excluded. Figure \ref{fig:sim_discrep} shows the inferred discrepancy under the elastic and standard approaches. Under the standard approach, the timing shift is captured by a sinusoidal discrepancy in amplitude, which limits the amount of information this part of the calibration can provide. On the other hand, the elastic approach finds a timing shift of 0.2 (the correct amount).

\section{Dynamic Material Model Calibration}\label{sec:results}

We demonstrate the application of the elastic functional Bayesian calibration method to two real world applications, a tantalum equation of state calibration and an aluminum material strength calibration. 

\subsection{Tantalum Z-Machine}
\label{sec:z-machine}
In this application, we seek to calibrate the equation of state (EoS) of tantalum (Ta) with data generated from pulsed magnetic fields \citep{brown2014}.
We are seeking to estimate parameters describing the compressibility (relationship between pressure and density) to understand better how materials compress to extreme pressures.
Ta is an ideal material for this study as it is able to remain in its initial crystal structure to pressures up to 10 million times standard atmospheric pressure \citep{soderlind:1998}.
A description of the Ta experiments is shown in Figure~\ref{fig:ta_data2}.
The experiments were conducted using Sandia National Laboratories' Z-machine, which is a pulsed power drive that can deliver massive electric currents over short time scales. 
These currents were forced to flow along an aluminum panel, producing a large magnetic pressure which drives a time-dependent stress wave (impulse) into the system. 
Ta samples and transparent lithium fluoride (LiF) windows were glued to the panel such that the stress wave propagates sequentially through each of these materials.

The material properties are modeled using a physically motivated form given by \cite{vinet:1989}.
This form describes the pressure-density $(P-\rho)$ response as 
\[P(\rho)=3B_0\left(\frac{1-\eta}{\eta^2}\right)\exp\left(\frac{3}{2}(B'_0-1)(1-\eta)\right),\]
where $\eta=\sqrt[3]{\rho_0/\rho}$, $\rho_0$ is the initial density and $B_0$ and $B
'_0$ are the bulk modulus and its pressure derivative at ambient conditions. 
From the computer experiment perspective, we will work with 6 inputs (3 EoS parameters ($\rho,B_0,B'_0$) and 3 experiment-specific parameters) and output velocity curves on a grid of 100 equidistant time points.  As described in \citep{brown2014}, the main goal is to provide inferences on the 3 EoS parameters with uncertainty quantification and to propagate these inferences to the Vinet model.

\begin{figure}[htbp]
	\centering
	\includegraphics[width=.85\textwidth]{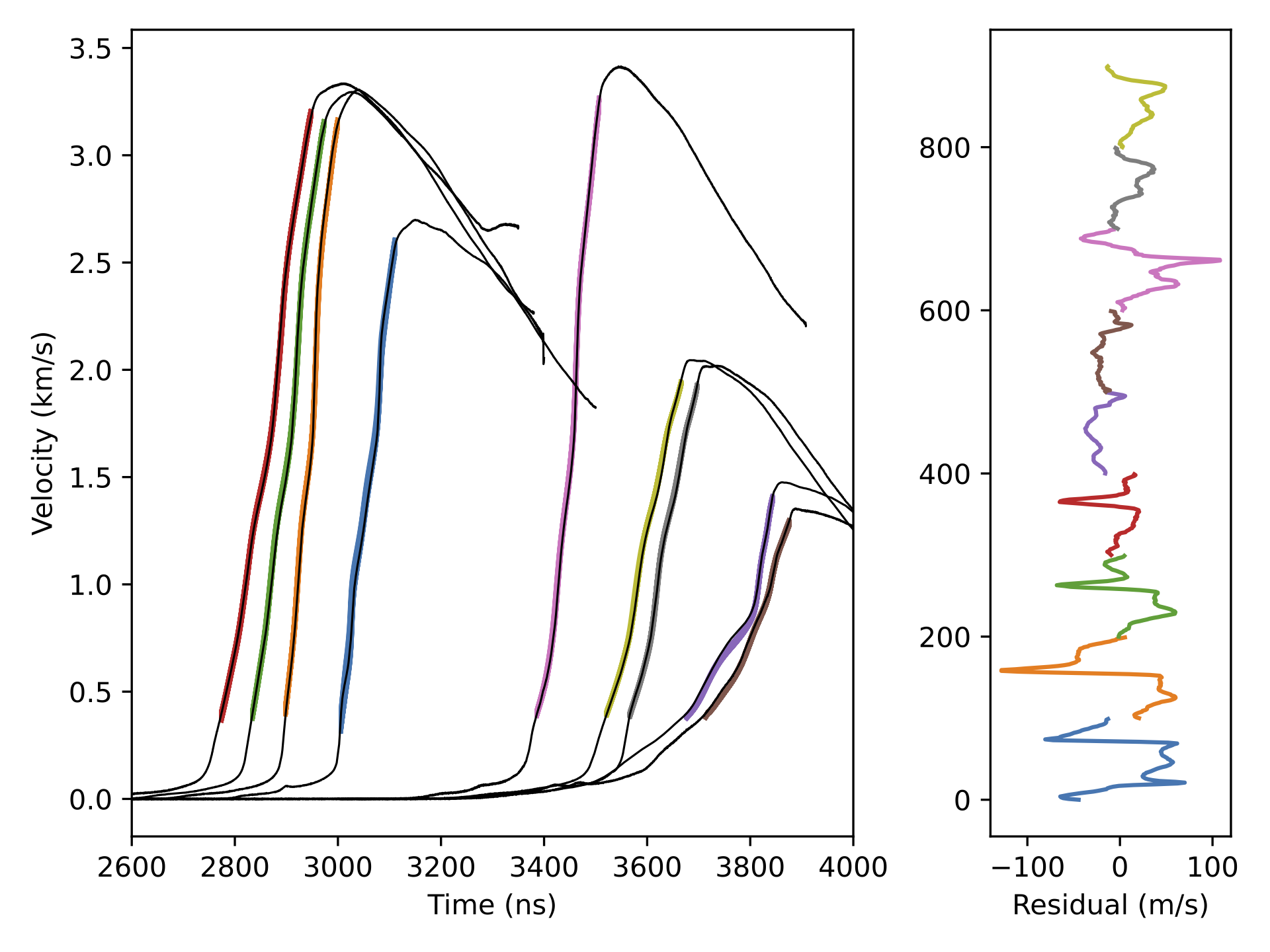}
	\caption{(Left) Experiment velocities of Ta shown in black compared with 95\% prediction intervals from the elastic functional Bayesian calibration. (Right). Corresponding color-coded residuals (difference between experiment and calibration mean) for each experiment}
	\label{fig:ta_calibration}
\end{figure}

As was done in the simulated example, the computer model output was aligned to the experimental data, and an emulator was fitted to the aligned computer model and the corresponding shooting vectors. We then performed a modular elastic Bayesian model calibration. There are a total of 9 experiments.
Figure~\ref{fig:ta_calibration} presents the elastic BMC results for the Ta experiments. 
The black curves shown in the figure correspond to the experimental velocity curves for the 9 experiments.
The shaded colored regions are the 95\% prediction intervals that result from the elastic functional Bayesian calibration. The prediction intervals exhibit good agreement with each experimental curve. The residuals defined as the difference between the experimental data and the calibrated predicted mean, are shown in the right panel of Figure~\ref{fig:ta_calibration}. Each of the residuals are color coded to the corresponding experiment from the left panel. The full functional approach has tighter coverage of the experimental curves and smaller residual values compared to the resulting predictions and residuals that stem from the approach of \cite{brown2018}.

\begin{figure}[htbp]
	\centering
	\includegraphics[width=.6\textwidth]{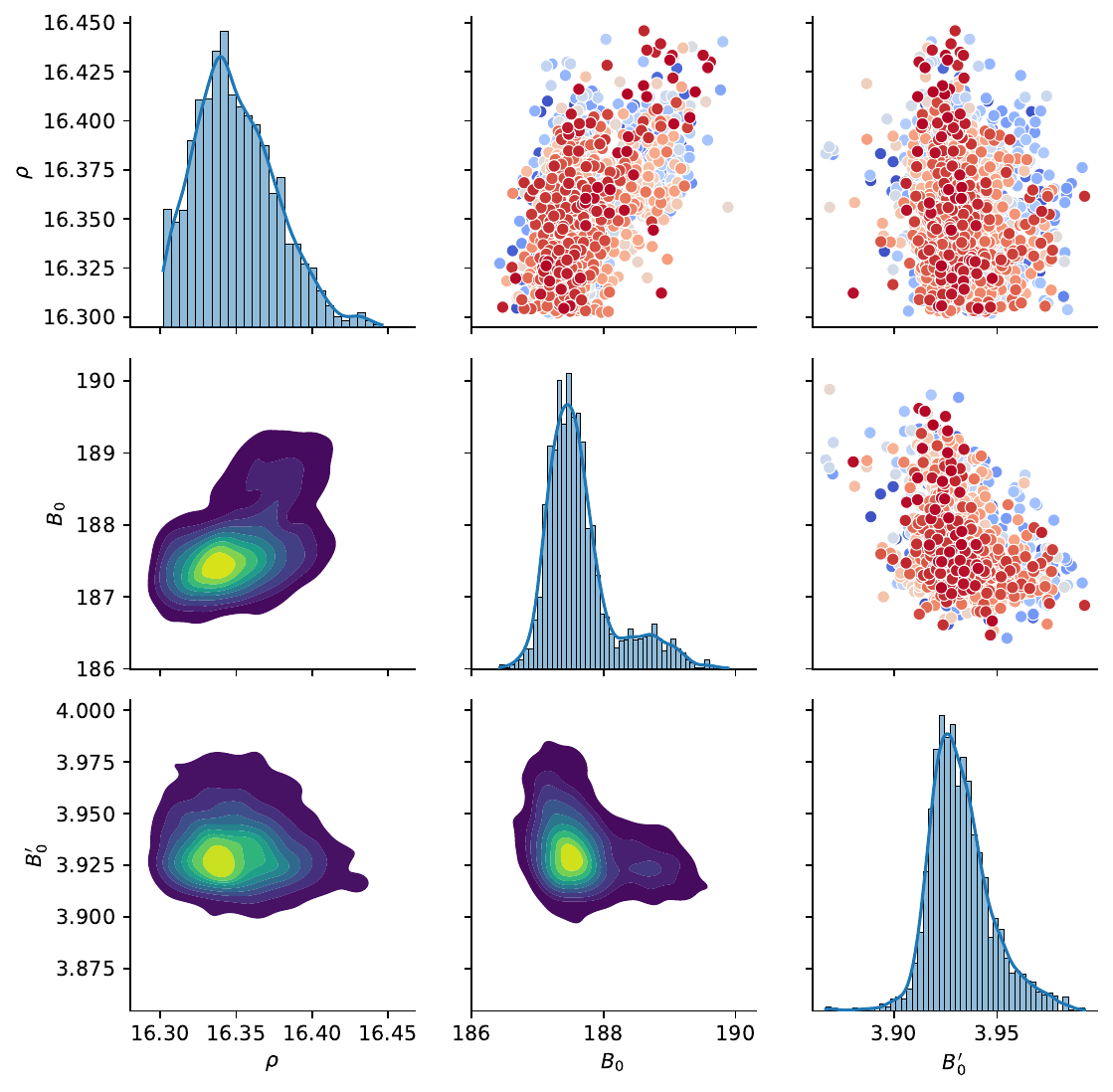}
	\caption{Pairs plot of the posterior densities of EoS parameters for Tantalum}
	\label{fig:ta_pairplot}
\end{figure}

\begin{figure}[htbp]
	\centering
	\includegraphics[width=.6\textwidth]{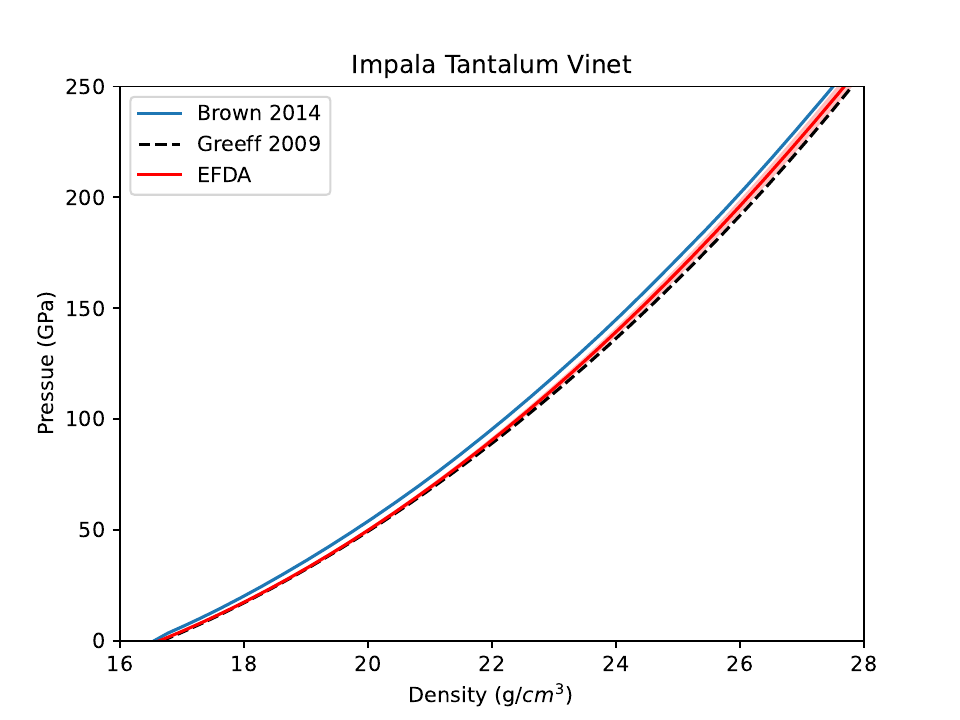}
	\caption{Calibrated Ta material response to the Vinet model compared with the analytic results in \cite{brown2014} and the theoretical calculations in \cite{greef:2009}.}
	\label{fig:vinet}
\end{figure}

Furthermore, Figure~\ref{fig:ta_pairplot} presents a pairwise plot of the samples from the posterior distribution of the calibrated EoS parameters for Ta with the elastic approach. The posterior distributions are well concentrated within the parameter space  and the corresponding  posterior medians compare to those found in \cite{brown2018}, with a reduced uncertainty around the same parameter values. It should be remarked that the elastic FDA calibration approach does not involve any likelihood scaling as in \cite{brown2018}.
To assess the calibration further, we compared the posterior estimates of the EoS to those reported in the literature for the Vinet model.
Figure~\ref{fig:vinet} presents a plot of the pressure versus density Vinet curve, with a 95\% credible interval generated from the elastic calibration shown in red. 
The blue curve corresponds to the loading path determined analytically by using the standard techniques in the dynamic materials community as described in \citep{brown2014}.
The dashed curve is the state of the art theoretical calculation given by \cite{greef:2009}, which has been used to simulate these types of Ta experiments.
The posterior estimate of the curve from the elastic calibration lies just above the theoretical calculation and below the analytical result, which corresponds with what is expected when using the analytic analysis \citep{kraus:2016}. The estimates of the physical parameters resulting from the elastic BMC are similar to those from previous work.

\subsection{Flyer Plate Impact}
Material strength characterizes how a material temporarily or permanently deforms as it experiences pressure. 
 This is of interest in various areas of science and engineering, with applications in the aerospace, medical, and automotive industries \citep{gray2005predicting}.  
 When experimentation is difficult, material strength models are an important tool for predicting how a material will react to pressure.  
One such model is the Johnson-Cook strength model \citep{johnson1983}, for which the model behavior is dictated by a relatively small collection of material-specific physical parameters.  To set these parameters, scientists and engineers rely on experimentation, hence calibrating the material strength model parameters to experimental measurements. 

\begin{figure}[htbp]
     \centering
     \begin{subfigure}[b]{0.7\textwidth}
         \centering
          \includegraphics[width=\textwidth]{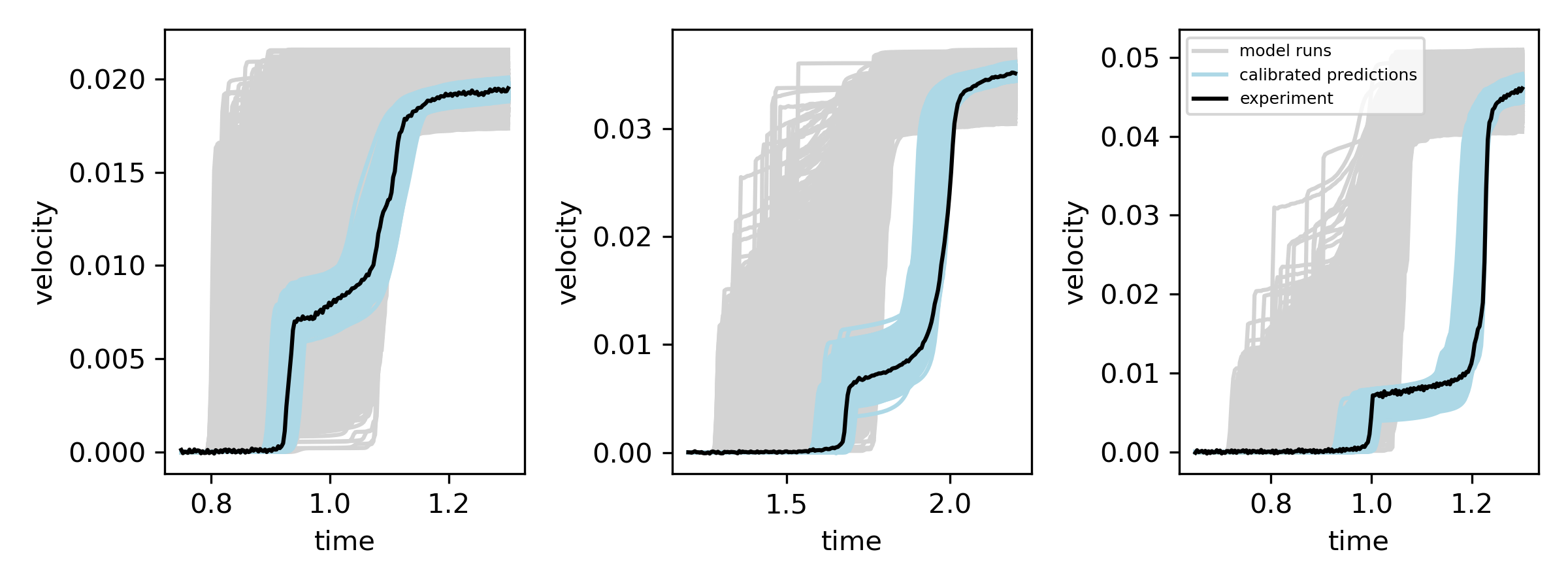}
         \caption{Original Data and 1000 simulated velocimetry curves, with calibrated posterior predictions.}
         \label{fig:flyer_original}
     \end{subfigure}
          \hfill
           \begin{subfigure}[b]{0.7\textwidth}
         \centering
          \includegraphics[width=\textwidth]{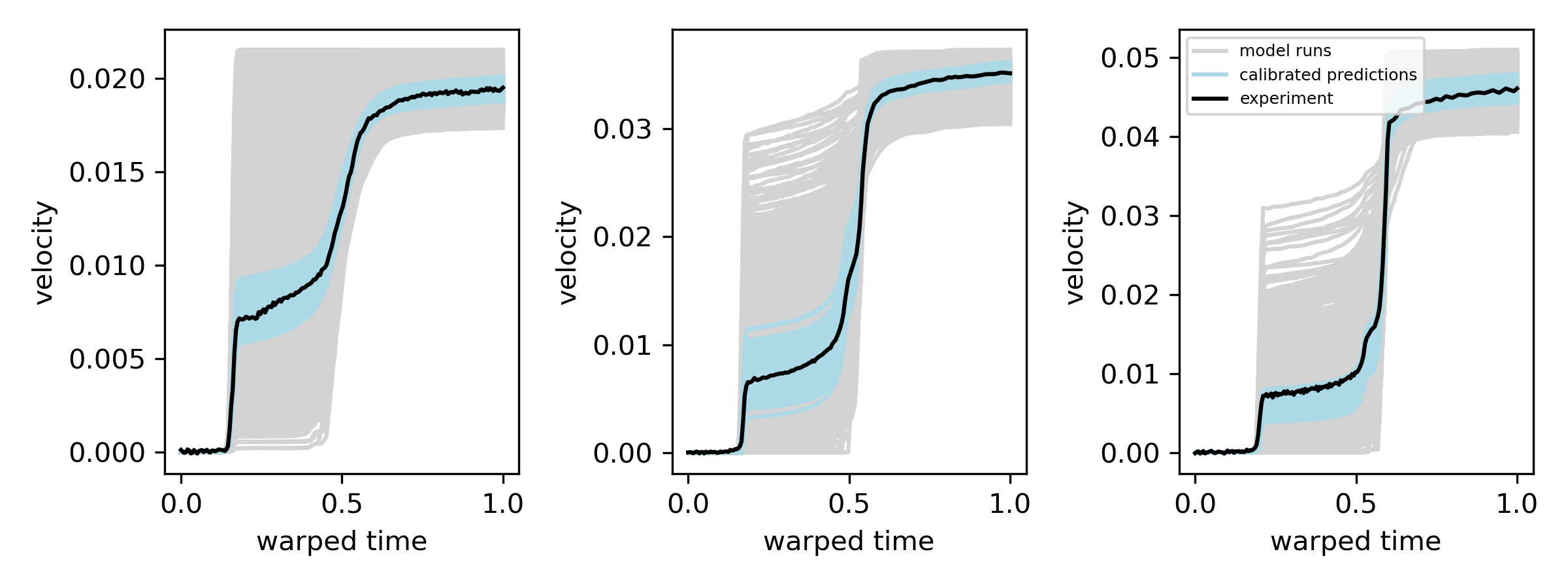}
         \caption{Aligned measurement and simulation curves, with calibrated posterior predictions.}
         \label{fig:flyer_calibration3}
     \end{subfigure}
         \hfill
         \begin{subfigure}[b]{0.7\textwidth}
         \centering
         \includegraphics[width=\textwidth]{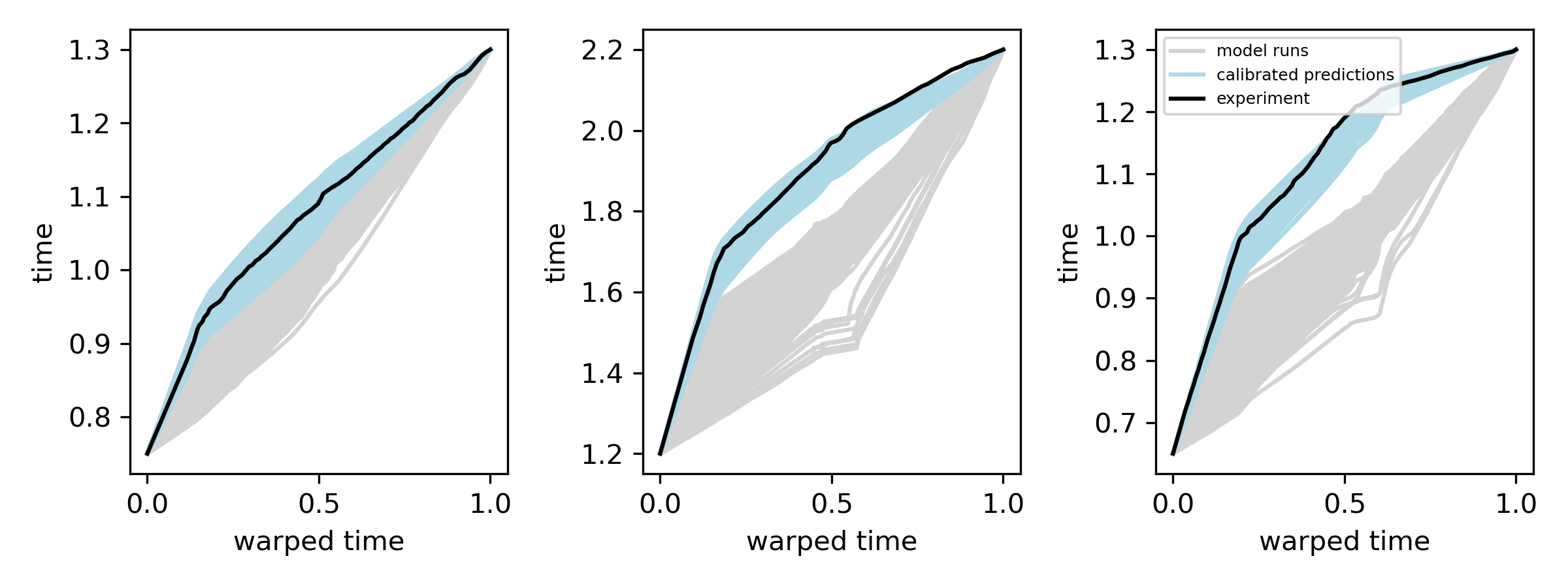}
         \caption{Warping functions for measurement and simulation curves, with calibrated posterior predictions.}
         \label{fig:flyer_calibration2}
     \end{subfigure}
     \begin{subfigure}[b]{0.7\textwidth}
         \centering
         \includegraphics[width=\textwidth]{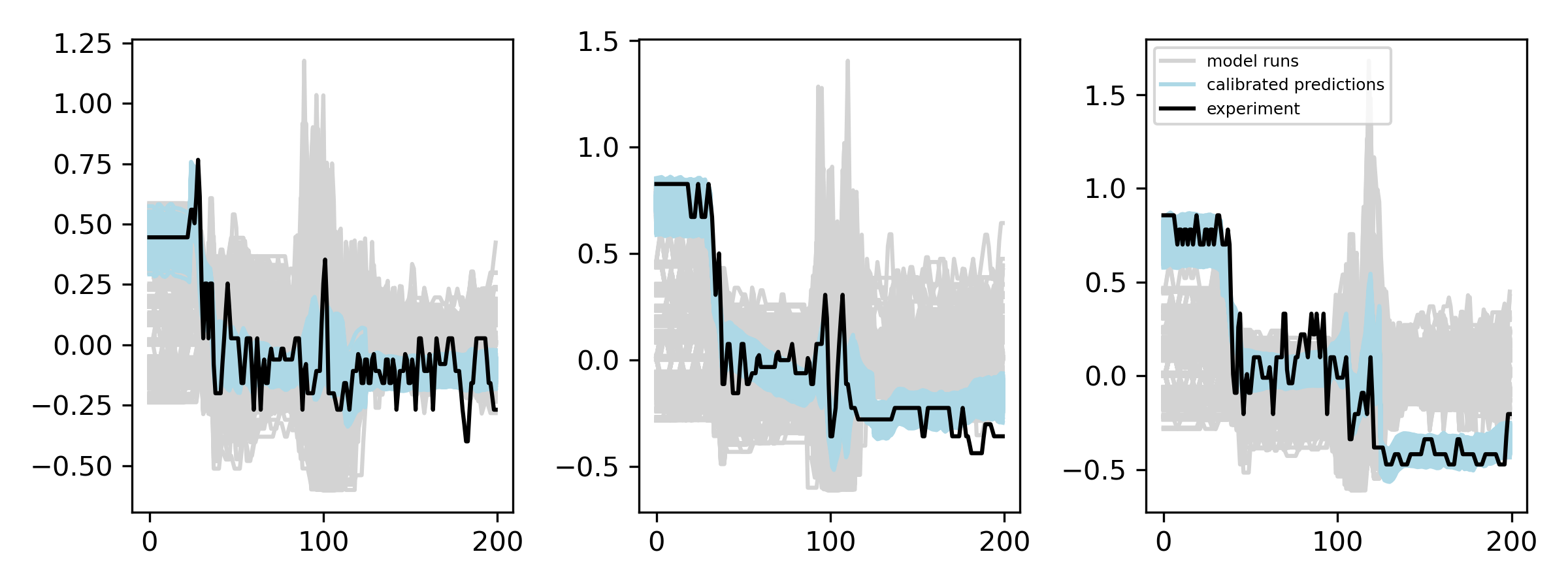}
         \caption{Shooting vectors for measurement and simulation curves, with calibrated posterior predictions.}
         \label{fig:flyer_calibration1}
     \end{subfigure}
     \caption{Posterior predictions from the elastic BMC for the flyer data example for the original data, the aligned curves, the warping functions and the shooting vectors.}
     \label{fig:flyer_calibration}
\end{figure}

In this application, we consider the calibration of the Johnson-Cook material strength model for an aluminum alloy using a set of plate impact experiments. 
\cite{walters2018bayesian} performed a Bayesian model calibration using plate impact experiments by reducing the velocimetry curves to a small set of features considered to be important to strength. We show how elastic Bayesian model calibration relies on the entire curve without using human-intensive feature engineering. Plate impact experiments achieve high pressure on a material sample by shooting a flyer at high velocity into the sample (plate).  Lasers measure how the free surface of the sample moves as the shock wave moves through it.  
The result is a trace of the velocity of the free surface of the sample over time, called a velocimetry curve.  
From the computer experiment perspective, we will work with 11 inputs: 5 Johnson-Cook parameters plus 6 experiment-specific parameters (two for each experiment) and the output velocimetry curves on a grid of 200 time points.  
Figure~\ref{fig:flyer_original}  shows measured velocimetry curves along with 1000 simulated velocimetry curves using different settings of the Johnson-Cook model and for three experiments.  To obtain these simulations, the Johnson-Cook model is used within a larger hydrodynamics code that is expensive to evaluate. In this example we do not use the observations as the alignment reference, and instead use the first model run. Figure~\ref{fig:flyer_calibration3} shows the aligned functions, Figure~\ref{fig:flyer_calibration2} shows warping functions, and Figure~\ref{fig:flyer_calibration1} shows the shooting vectors corresponding to the measured and simulated curves. The misalignment of the experimental curves to the simulated is clear for the second and third experiments. The two right plots of Figure~\ref{fig:flyer_original} show the experimental data to the right of all of the simulations, This is reflected in the warping functions in the two right plots of Figure~\ref{fig:flyer_calibration2}, which push the experimental timings later than any of the simulations. In the shooting vectors of Figure~\ref{fig:flyer_calibration1}, this is reflected with large positive values early in the vector and large negative values lat in the vector, with varying patterns in the middle of the vector.

\begin{figure}
	\centering
	\includegraphics[width=.7\textwidth]{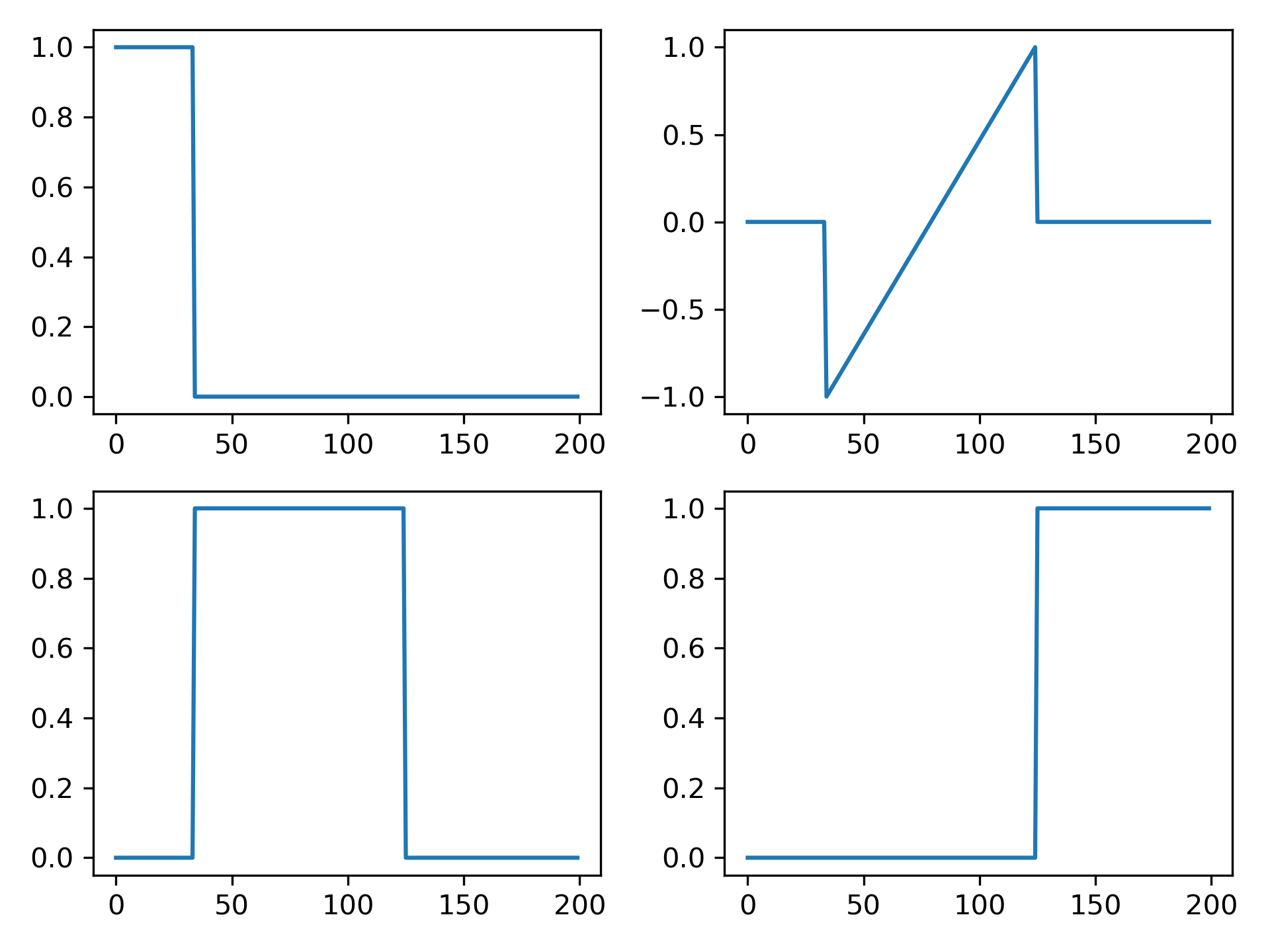}
	\caption{Basis functions used to capture a time shift discrepancy in shooting vector space for the flyer plate example.}
	\label{fig:flyer_discrep}
\end{figure}

\begin{figure}
	\centering
	\includegraphics[width=\textwidth]{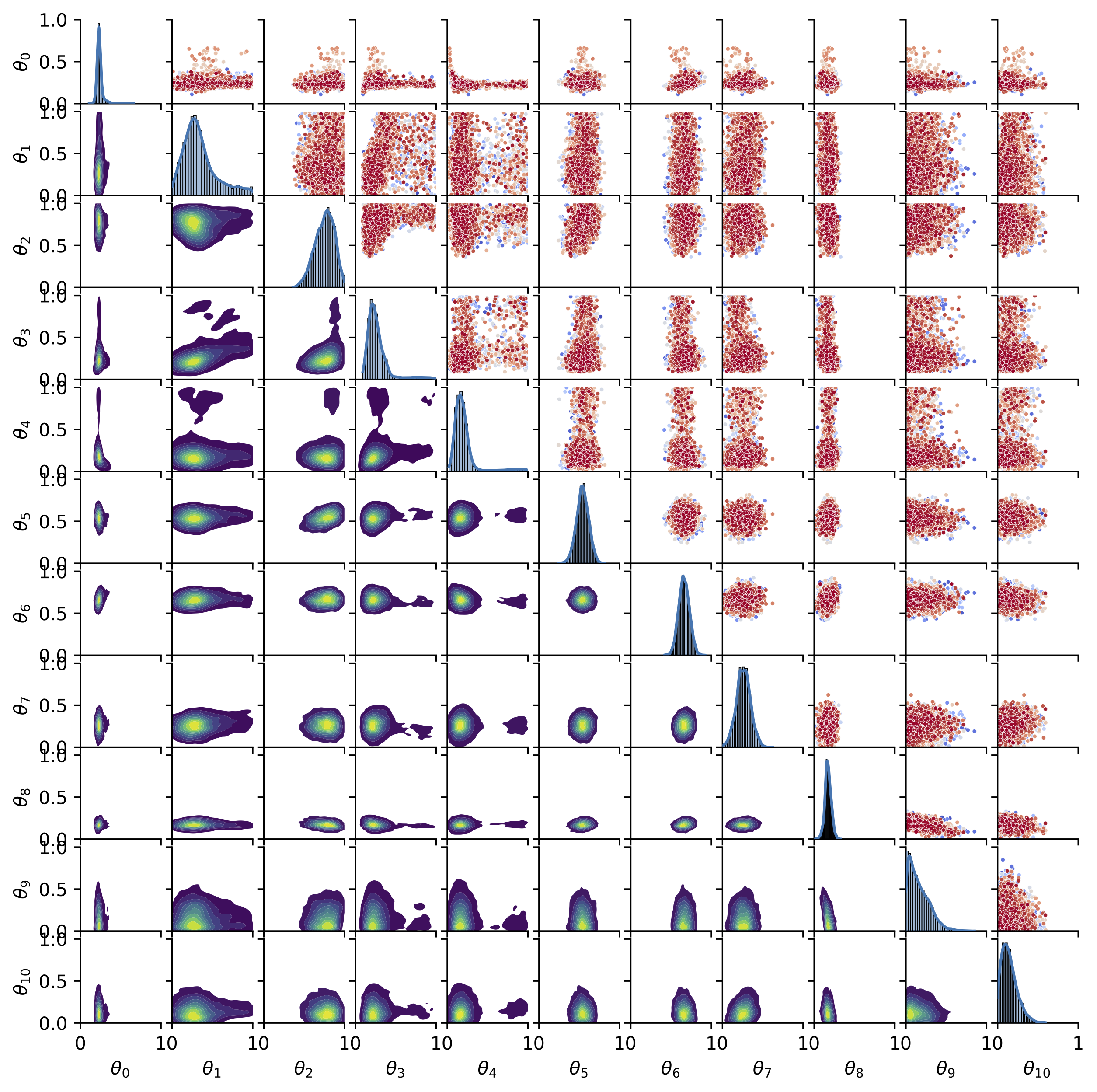}
	\caption{Pairwise plot and marginal densities for the posterior distribution of the 11 calibrated parameters for the flyer plate experiment.}
	\label{fig:flyer_pairplot}
\end{figure}

\begin{figure}[htbp]
     \centering
     \begin{subfigure}[b]{0.49\textwidth}
         \centering
         \includegraphics[width=\textwidth]{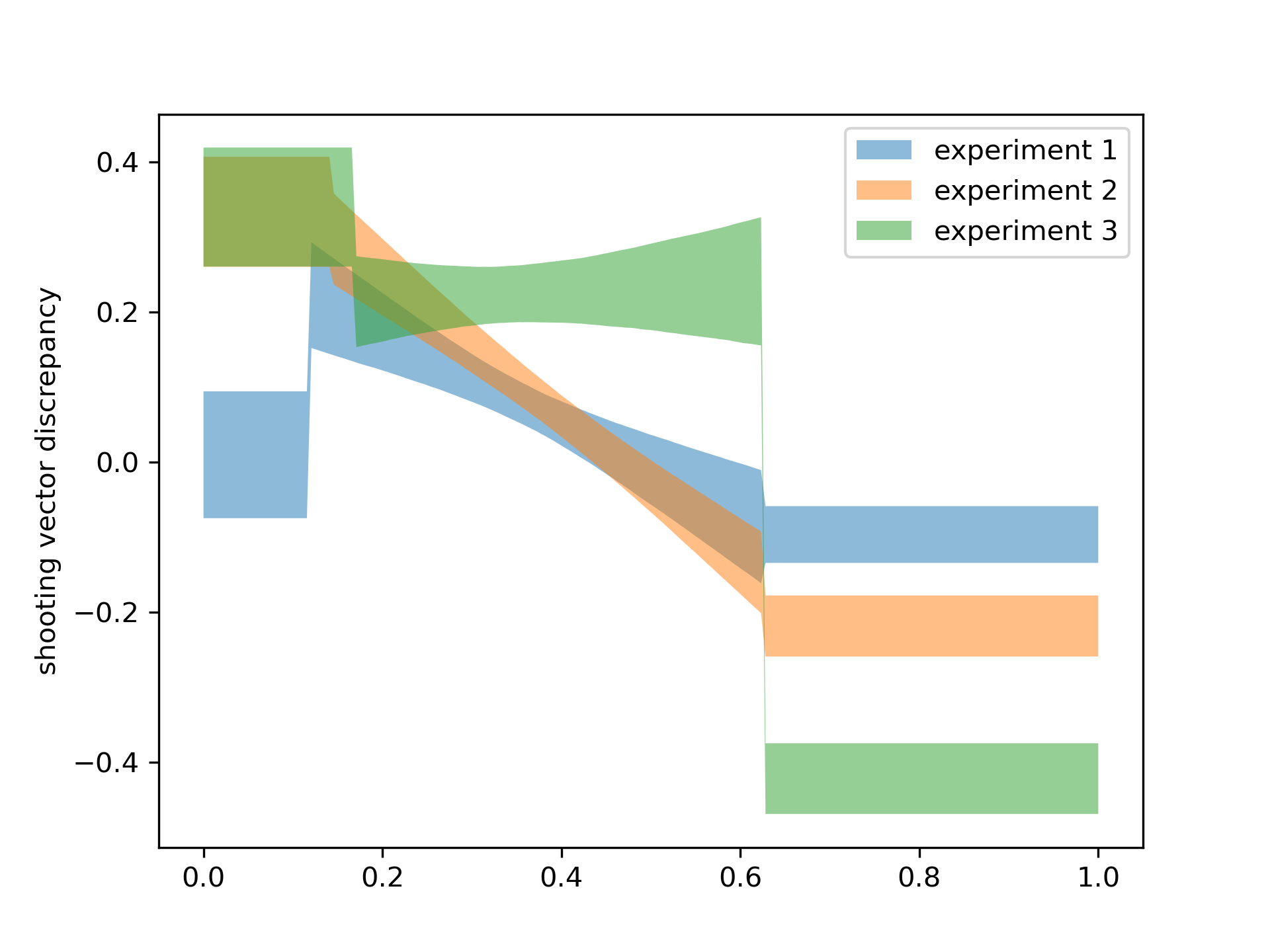}
         \caption{Discrepancy in shooting vector space.}
         \label{fig:flyer_vv_discrep}
     \end{subfigure}
     \begin{subfigure}[b]{0.49\textwidth}
         \centering
         \includegraphics[width=\textwidth]{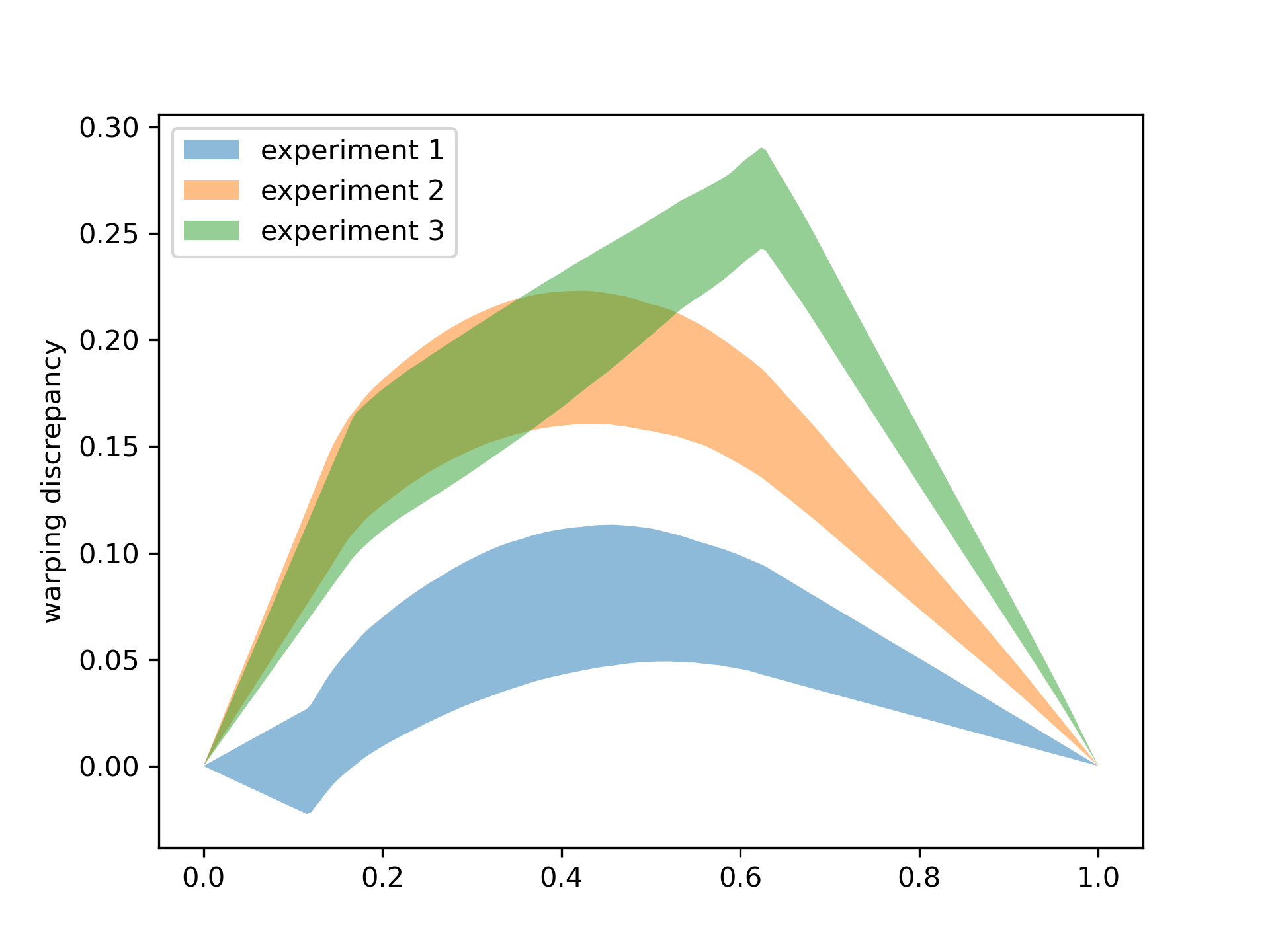}
        \caption{Discrepancy (additive) in warping function space.}
         \label{fig:flyer_gam_discrep}
     \end{subfigure}
     \caption{Estimated posterior discrepancy per experiment for the Flyer plate impact calibration.}
     \label{fig:flyer_discrep_post}
\end{figure}

We use the aligned data and shooting vectors from the 1000 simulations to build an emulator and perform the elastic calibration. 
 To allow for a time shift discrepancy for experiments 2 and 3, we use three piece-wise constant and one piece-wise linear basis functions in shooting vector space as shown in Figure~\ref{fig:flyer_discrep}.  More specifically, we parameterize $\bm\delta_v(\bm x_i) = \bm D \bm \beta(\bm x_i)$ where $\bm D$ is the $200 \times 4$ matrix of basis functions and $\bm \beta(\bm x_i)$ is a vector of $4$ basis coefficients that can vary by experiment.  In the approach of \cite{higdon2008computer}, $ \beta_k(\cdot)$ would be assigned a Gaussian process prior so that the discrepancy could be predicted for a new experiment with settings $\bm x^*$.  In our case, we are not interested in predicting the discrepancy at new experimental settings, so we specify a standard normal prior for each $\beta_k(\bm x_i)$.  This indicates that we want to allow for a time shift, but would prefer to have make $\beta_k(\bm x_i)=0$ if possible.  The prior variance of $\beta_k(\bm x_i)$ can be modified in order to favor more or less discrepancy.  The form of the basis functions in Figure~\ref{fig:flyer_discrep} is carefully chosen to be constant in domains where the aligned data are roughly constant and a linear function in the region where the aligned data are most active. Specifically, we think of the aligned curves as having three parts: (1) a nearly constant region at the beginning; (2) a quick jump in velocity, a plateau, and another quick jump; and (3) a final plateau. We choose our basis functions such that we allow for constant discrepancy shifts in the shooting vectors where in the (1) and (3) regions, and linear shifts in the (2) region. We largely arrived at this combination by trial and error made sensible in hindsight. Recall that the transformation of shooting vector to warping function involves the exponential map and an integral. Portions of the shooting vector that are positive indicate faster-than-identity warping, while negative values indicate slower-than-identity warping. The warping functions are constrained to start at 0 and end at 1, so the constant shifts in shooting vectors at the first and last plateaus allow for more extreme change in timing there, where it does not have much effect, to compensate for the linear portion of the shooting vector discrepancy, which has the more noticeable effect.

Figure~\ref{fig:flyer_calibration} presents the  posterior predictive samples of the velocimetry curves after calibration shown in blue. In a similar way, Figures~\ref{fig:flyer_calibration3}, \ref{fig:flyer_calibration2} and \ref{fig:flyer_calibration1} show the posterior predictive samples of the aligned curves, warping functions and shooting vectors respectively. 
In all of these cases, the predictive samples cover the experimental data well. 
Furthermore, Figure~\ref{fig:flyer_pairplot} presents a pairs plot of the posterior distribution of the 11 calibrated parameters with diagonal elements representing marginal distributions while the lower diagonal shows bivariate contours and the upper diagonal shows pair plots of the samples. Compared to \cite{walters2018bayesian}, we are able to use the entirety of the functional data while their approach used a few hand selected features. 

Figure~\ref{fig:flyer_vv_discrep} shows the posterior distribution of $\bm\delta_v(\bm x_1)$, $\bm\delta_v(\bm x_2)$, and $\bm\delta_v(\bm x_3)$, the shooting vector discrepancies.  Figure~\ref{fig:flyer_gam_discrep} shows the difference between two sets of warping functions, those that include the discrepancy in Figure~\ref{fig:flyer_vv_discrep} and those that do not, which illustrates the type of shift that these discrepancies induce.

The code to reproduce this analysis is available at \url{https://github.com/lanl/impala/blob/master/examples/shpb-flyer-Al5083_pooled.py}.

\section{Conclusion}\label{sec:discussion}
Model calibration for functional responses is typically more difficult than in more traditional settings.  
Adjusting input parameters and discrepancy so model output matches experimental output is seldom trivial, but can be even more challenging for functional data when altering input parameters results in amplitude and phase variability in the responses.  
Traditional methods ignore these aspects which are unique to models with functional responses, putting them at risk for higher levels of bias and lower levels of efficiency.  
In this paper we develop methods to handle amplitude and phase variability in a systematic way, resulting in more efficient and more accurate estimates of the model parameters.  
The improvements in these estimators is achieved through better handling of the functional responses as opposed to collecting more data.

The elastic Bayesian model calibration procedure presented here uses information from both the amplitude and phase space to calibrate the parameters, in contrast with 
more common calibration methods which would only look at the amplitude space.  
The benefits of this approach are demonstrated on a toy problem where amplitude and phase variability are present.  
The toy problem shows the elastic Bayesian model calibration approach results in superior estimation of the model parameters and predicted functional responses.  
Information about the amount of warping necessary to align the functions provides an additional indirect benefit, as it provides valuable insight into the model’s input parameters.  
Standard methods which do not handle the phase and amplitude variability separately, may result in predicted functional responses which do not fit the data well and suggest much larger uncertainty in calibration parameters.  
These results make a strong case for the elastic modeling approach, as most functional data have some misalignment and phase discrepancy.  

Applying the method to the Z-machine data and flyer-plate data generated similar results to previous studies without the need of likelihood scaling as in \cite{brown2018} or feature selection as in \cite{walters2018bayesian}.  
The elastic Bayesian calibration model yields credible intervals with good agreement of the experimental data using a principled approach.  
From a theoretical perspective, the elastic Bayesian calibration approach is satisfying in that it treats the model’s output as functional throughout the calibration procedure.  

Potential future work may investigate some of the additional modeling assumptions made in the elastic calibration approach, such as the independence assumptions in the likelihood and the treatment of the warping decomposition uncertainty.

\section*{Acknowledgment}
This paper describes objective technical results and analysis. Any subjective views or opinions that might be expressed in the paper do not necessarily represent the views of the U.S. Department of Energy or the United States Government. This work was supported by the NA-22 program, the Advanced Simulation and Computing program, and Laboratory Directed Research and Development program at Los Alamos National Laboratory and Sandia National Laboratories. Sandia National Laboratories is a multi-mission laboratory managed and operated by National Technology and Engineering Solutions of Sandia, LLC, a wholly owned subsidiary of Honeywell International, Inc., for the U.S. Department of Energy's National Nuclear Security Administration under contract DE-NA0003525.

\bibliographystyle{plainnat}
\bibliography{bibl}
		
\end{document}